\documentclass[12pt]{article}
\usepackage{amsmath}
\usepackage{amsfonts}

\input{tcilatex}
\begin{document}

\title{Explicit mass renormalization and consistent derivation of radiative
response of classical electron}
\author{Anatoli Babin \\
Department of Mathematics, University of California, \\
Irvine, U.S.A.}
\date{}

\begin{abstract}
The radiative response of the classical electron is commonly described by
the Lorentz-Abraham-Dirac (LAD)\ equation. Dirac's derivation of this
equation is based on energy and momentum conservation laws and on
regularization of the field singularities and infinite energies of the point
charge by subtraction of certain quantities: "We ... shall try to get over
difficulties associated with the infinite energy of the process by a process
of direct omission or subtraction of unwanted terms". To substantiate
Dirac's approach and clarify the mass renormalization, we introduce the
point charge as a limit of extended charges contracting to a point; the
fulfillment of conservation laws follows from the relativistic covariant
Lagrangian formulation of the problem. We derive the relativistic point
charge dynamics described by the LAD equation from the extended charge
dynamics in a localization limit by a method which can be viewed as a
refinement of Dirac's approach in the spirit of Ehrenfest theorem. The model
exhibits the mass renormalization as the cancellation of Coulomb energy with
the Poincar\'{e} cohesive energy. The value of the renormalized mass is not
postulated as an arbitrary constant, but is explicitly calculated. The
analysis demonstrates that the local energy-momentum conservation laws yield
dynamics of a point charge which involves three constants: mass, charge and
radiative response coefficient $\theta $. The value of $\theta $ depends on
the composition of the adjacent potential which generates Poincar\'{e}
forces. The classical value of the radiative response coefficient is singled
out by the global requirement that the adjacent potential does not affect
the radiated energy balance and affects only the local energy balance
involved in the renormalization.
\end{abstract}

\maketitle




\section{Introduction}

The dynamics of the classical electron which involves its radiative response
is described by the Lorentz-Abraham-Dirac (LAD)\ equation which has the
form, \cite{Dirac CE}, \cite{Kosyakov}, \cite{LandauLif F}, \cite{Panofsky
Phillips}, \cite{Spohn}, \cite{Thirring}: 
\begin{equation}
m\dot{v}_{\mu }=ev_{\nu }F_{\mathrm{ex}\mu }^{\nu }+\frac{2}{3}e^{2}\ddot{v}%
_{\mu }+\frac{2}{3}e^{2}\left( \dot{v}\dot{v}\right) v_{\mu }.  \label{ALDeq}
\end{equation}%
Here the covariant notation \ is used, $v_{\mu }$ is 4-velocity of a charged
point, $f_{\mathrm{ex}\mu }=ev_{\nu }F_{\mathrm{ex}\mu }^{\nu }$ is the
Lorentz force generated by the external field $F_{\mathrm{ex}}^{\mu \nu }$, $%
\dot{v}_{\mu }=\partial _{s}v_{\mu }$ and $\left( \cdot \cdot \right) $ is
the 4-product in Minkowski space: 
\begin{equation}
\left( vw\right) =v_{\nu }w^{\nu }=g_{\mu \nu }v_{\mu }w_{\nu
}=v_{0}w_{0}-v_{1}w_{1}-v_{2}w_{2}-v_{3}w_{3}=v_{0}w_{0}-\mathbf{v}\cdot 
\mathbf{w};  \label{vw}
\end{equation}%
we use the units in which the speed of light $\mathrm{c}=1$ and the
summation convention. The LAD equation does not describe the magnetic moment
of the electron, for generalizations in this direction see \cite{Appel
Kiessling}, \cite{Nodvik}.\ The radiative response of an electron was
originally derived by Abraham and Lorentz \ from the analysis of the
Lorentz-Abraham \ model, see \cite{Jackson}, \cite{Rohrlich}, \cite{Spohn}, 
\cite{Yaghjian}, \cite{Yaghjian1}. A relativistic treatment of the
Lorentz-Abraham model meets with difficulties, \cite{Jackson}, \cite%
{Panofsky Phillips}, \cite{Pearle1}, \cite{Rohrlich}, \cite{Spohn}, \cite%
{Thirring}, \cite{Yaghjian}, \cite{Yaghjian1}. The relativistic derivation
of the radiative response based on the analysis of the energy and momentum
conservation laws and on mass renormalization is due to Dirac \cite{Dirac CE}
and now is often used, \cite{LandauLif F}, \cite{Panofsky Phillips}, \cite%
{Thirring}. It is well-known that the derivation of the LAD equation and the
equation itself is not without difficulties, \cite{Kosyakov}, \cite{Panofsky
Phillips}, \cite{Rohrlich}, \cite{Spohn}. \ Sometimes the source of \ the
difficulties is attributed to the mass renormalization. The mass
renormalization is described in \cite[Sec. 75]{LandauLif F} as follows:
"When in the equation of motion we write a finite mass for the charge, then
in doing this we essentially assign to it a formally infinite negative
"intrinsic mass" of nonelectromagnetic origin, which together with the
electromagnetic mass should result in a finite mass for the particle. Since,
however, the subtraction of one infinity from another is not an entirely
correct mathematical operation, this leads to a series of further
difficulties". One could think that since a constant can be added to the
total energy, then the infinite self-energy is not a problem. But one has to
fulfil the mass renormalization in dynamical regimes with acceleration,
where external forces may change the energy of the charge, and it cannot be
considered constant. In addition, in a relativistic setting one cannot treat
the energy as an independent quantity and must consider the energy-momentum
4-vector. Therefore, it is not clear that the mass renormalization in
non-trivial dynamical regimes with self-interaction is "an entirely correct
mathematical operation". An implicit assumption that the removal of
infinities from a model is a "surgery" which results only in a change of the
renormalized parameter of the model and does not have any side effects is
not necessarily true. But we show below that though side effects exist, they
are controllable.

A natural way to deal with the difficulties in the treatment of infinities
is to introduce an extended charge and then apply to it Dirac's analysis to
find the limit dynamics in the point localization limit as its size tends to
zero. But there are difficulties in this approach. Namely, \cite[Sec. 8.4]%
{Thirring}: "Unfortunately, it is not easy to obtain a theory in this way
that has a local conservation of energy and momentum". Nevertheless, we
follow this path; a different approach to the dynamics of a point charge in
EM field, which does not use renormalization and is based on
Maxwell-Born-Infeld equations, is developed in \cite{Kiessling 1}. \ The
fulfillment of the local energy-momentum conservation laws up to the point
limit constitutes an important part of Dirac's method: "The usual derivation
of the stress-tensor is valid only for continuous charge distributions and
we are here using it for point charges. This involves adopting as a
fundamental assumption the point of view that energy and momentum are
localized in the field in accordance with Maxwell's and Pointing's ideas", 
\cite[p. 152]{Dirac CE}.

Therefore, to obtain the point charge localization limit so that Dirac's
analysis still remains applicable, we model an extended charge in the
framework of a relativistic invariant field theory which produces
corresponding conservation laws. Such an approach inevitably has to meet
with certain difficulties. An extended charge, in a contrast to the point
charge, is subjected to Coulomb repulsive forces, and Poincar\'{e} cohesive
forces should be present to provide stability of the charge. Though the
Coulomb energy of an extended charge is finite, the energy tends to infinity
in the point limit. The "mass renormalization", which is an important
component of Dirac's method, in a self-consistent model should be derived
from the model and not postulated. The value of the renormalized mass of the
charge, which is simply prescribed for a point charge, must be consistently
derived from the analysis of the model. Such a model would demonstrate that 
\emph{Dirac's method, including the mass renormalization, is fundamentally
correct}. Another advantage of constructing an explicit model is a
possibility to answer the following important questions: \emph{Does the
limit dynamics of the point charge depend on the way the point charge is
obtained from an extended charge? Is there an effect of the cohesive forces
on the charge dynamics which is not described by the renormalized value of
mass?\ }

Presentation and analysis of a model of a point charge which satisfies the
above requirements is the subject of this paper. Our purpose is to show that
all components of Dirac's approach, including the mass renormalization, can
be consistently and rigorously realized. \emph{We introduce a classical
relativistically covariant model of a charge with fulfillment of the energy
and momentum conservation laws which allows asymptotic localization of a
charge to a point and is simple enough, so we can find explicitly the limit
dynamics and study the effects of the mass renormalization.} The distributed
charge is described here not by the Lorentz-Abraham model or its
generalizations, \cite{Appel Kiessling}, \cite{Kiessling2}, \cite{Nodvik}, 
\cite{Pearle1}, \cite{Spohn}, \cite{Yaghjian}, but by a relativistic
covariant Lagrangian field theory as in \cite{BF8}, \cite{BF9}, where the
spatial distribution of the charge is not prescribed but rather is
determined as a solution of a field equation.

To balance destabilizing EM forces, we introduce in the model cohesive
forces of a non-electromagnetic origin which play the role of Poincar\'{e}
forces, they are introduced explicitly through an additional internal field
of a non-electromagnetic nature acting only on the charge itself; we call
the balancing field "adjacent field" of a charge. The adjacent field serves
two main purposes. First, it makes a distributed charge stable, providing
necessary Poincar\'{e} stresses \ which compensate for Coulomb \
self-repulsion. Second, it provides the charge energy renormalization. The
relevance of Poincar\'{e} stresses for dynamical properties of a charge in
the non-radiative case is well-known, \cite{Jackson}, \cite{LandauLif F}, 
\cite{Panofsky Phillips}, \cite{Rohrlich}, \cite{Thirring}, \cite{Yaghjian},
and is demonstrated in Poincar\'{e}-Schwinger model, \cite{Jackson}, \cite%
{Poincare}, \cite{Schwinger}. Dirac's derivation of the LAD equation uses
point charges from the very beginning, therefore Poincar\'{e} stresses do
not emerge, though the energy and momentum conservation laws are assumed to
be fulfilled and implicitly non-electromagnetic cohesive forces are taken
into account through the mass renormalization. In our model both the Poincar%
\'{e} stresses and the energy of non-electromagnetic origin required for the
mass renormalization originate from the adjacent field of the charge,
therefore the mass renormalization is intrinsically connected with the
Poincar\'{e} stresses.\emph{\ In the point localization limit the internal
structure of the extended charge and the adjacent field disappear, but not
without a trace: the value of the mass and the magnitude of the radiation
reaction in the LAD\ equation depend on the parameters which determine
dynamics of the extended charge and its adjacent field. }Note that the
introduction of Poincar\'{e} stresses based on a combination of the adjacent
field with the nonlinearity is not the only possible way to balance the EM
self-interaction for an extended charge in the field framework; it is
possible to use only a properly defined nonlinearity without use of an
adjacent field as in \cite{BF5}, but the analysis in the point limit meets
with technical difficulties. The model involving the adjacent field which we
use here allows for a detailed analysis in the point limit, and it is closer
to the original Dirac's approach. Namely, the energy-momentum tensor
generated by the adjacent field cancels out the principal singularities of
the energy-momentum tensor of the EM\ field exactly in the same way Dirac
removes the principal singularity of the EM field energy-momentum tensor
using subtraction of a regularizing field.

In the model which we study here, we are able to explicitly observe two
infinities -- the infinite energy of Coulomb field and the infinite energy
of \ the adjacent field, the infinities emerge as the size of the extended
charge tends to zero. Our analysis explicitly shows that these infinite
energies cancel out almost completely, and only a finite energy remains,
this finite part is shown to be responsible for the observable mass of the
charge. Therefore, the mass renormalization becomes a self-consistent,
explicit and quite regular procedure. Note that all the energies in question
are canonically defined in terms of corresponding Lagrangians, therefore the
internal energy and ultimately the mass of the charge is defined uniquely.
This is in a contrast with the usual mass renormalization where a certain
amount of energy is subtracted to obtain a finite energy; the subtracted
energy is interpreted as the energy of cohesive forces and the finite energy
is interpreted as the renormilized mass and its value is prescribed
arbitrarily. Our method of determination of the limit dynamics can be
considered as a refinement in the spirit of Ehrenfest theorem of Dirac's
approach \cite{Dirac CE}; this method was applied in \cite{BF8}, \cite{BF9}
to derive Einstein's formula in regimes with acceleration in the case where
there is no radiative response. Our derivation is explicit in the sense that
not only the rest mass is calculated, but also it describes all effects of
Poincar\'{e} \ cohesive forces relevant to the point charge dynamics, and in
particular demonstrates that their dynamical effect cannot be reduced to one
constant (the renormalized mass), but involves in addition another parameter
which controls the relative magnitude of \ the radiative response.

Now we briefly describe basic features of our method. \ The LAD equation for
a point charge motion is derived under the assumption of the asymptotic
localization of the energy of the charge. We do not prescribe a mass
distribution for the charge, in the process of derivation we obtain
Einstein's formula for the equivalence of the mass and energy which is not
postulated. The form of the equations of motion is uniquely determined by
the original assumptions in a contrast to the original Dirac's derivation
which allowed some freedom in the choice of the form of radiative reaction;
more complicated versions were rejected by Dirac \cite[p. 154]{Dirac CE}
since "they are all much more complicated, ... so one would hardly expect
them to apply to a simple thing like an electron". Remarkably, the
generalized LAD equation which we obtain has the form of the classical LAD
equation, but with one important difference. \emph{The radiation reaction
term has a constant coefficient }$\theta $\emph{\ which can take positive or
negative values:} 
\begin{equation}
m\dot{v}_{\mu }=ev_{\nu }F_{\mathrm{ex}\mu }^{\nu }+2\theta \left( \frac{2}{3%
}e^{2}\ddot{v}_{\mu }+\frac{2}{3}e^{2}\left( \dot{v}\dot{v}\right) v_{\mu
}\right) .  \label{vw1}
\end{equation}%
The dependence on $\theta $ demonstrates that if one considers a point
charge as a limit of contracting continuously distributed charges described
by a relativistic invariant Lagrangian field theory, the resulting point
dynamics exists, but depends on the way the distributed charges are
introduced. Dirac's original derivation produces LAD equation with the
specific value of the coefficient $\theta =1/2$, and this specific value can
be tracked down in the specific regularization of the EM field of the charge
obtained by subtracting the half-sum $\frac{1}{2}\left( F_{\mathrm{ret}%
}^{\mu \nu }+F_{\mathrm{adv}}^{\mu \nu }\right) $ of the advanced and
retarded fields of the charge, \cite[formula (13) p. 152]{Dirac CE}. This
regularization is an important part of Dirac's renormalization method \cite[%
p. 149]{Dirac CE}: "We shall retain Maxwell's theory to describe the field
up to the point-singularity which represents our electron and shall try to
get over difficulties associated with the infinite energy of the process by
a process of direct omission or subtraction of unwanted terms".

Our derivation is based on the following assumptions:

The elementary charge dynamics is described by a Lorentz and gauge invariant
Lagrangian which describes the system "charge-fields", involves the
following fields and has the following properties:

(i) The charge distribution field $\psi $ (it is a complex scalar in the
simplest case which we consider here, but not necessarily; for example it
could be a spinor).

(ii) The electromagnetic (EM) field $F^{\mu \nu }$ of the charge with the\
EM 4-potential $A^{\nu }$which is responsible for the EM interaction between
the charges, the EM field is governed by Maxwell's equations:%
\begin{equation}
\partial _{\mu }F^{\mu \nu }=4\pi J^{\nu }.  \label{fmunu}
\end{equation}

(iii) The adjacent field $\ F^{\mathrm{ad}\mu \nu }$ of the charge with the\
potential $A^{\mathrm{ad}\nu }$ of the charge, this field describes an
internal interaction of the charge with itself and balances the EM\
self-interaction of the charge. This field generates Poincar\'{e} cohesive
forces, it is also responsible for the mass renormalization. The adjacent
field is also governed by Maxwell's equations with the same source:%
\begin{equation}
\partial _{\mu }F^{\mathrm{ad}\mu \nu }=4\pi J^{\nu }.  \label{fadmunu}
\end{equation}

(iv) The source for the Maxwell equations for the EM field and the adjacent
field is the current $J^{\nu }$ which is determined by the charge
distribution field $\psi $.

(v) The field equation for $\psi $, in the simplest spinless case which we
consider here, has the form of \ the nonlinear Klein-Gordon (KG) equation%
\begin{equation}
\tilde{\partial}^{\mu }\tilde{\partial}^{\mu }\psi +\frac{m^{2}}{\hbar ^{2}}%
\psi +G^{\prime }\left( \psi ^{\ast }\psi \right) \psi =0;  \label{KG0}
\end{equation}%
it involves a nonlinearity $G^{\prime }=G_{a}^{\prime }$ which provides a
localization of the charge distribution characterized by a microscopically
small spatial scale $a$, $a\rightarrow 0.$

Note that Poincar\'{e} cohesive forces are required to balance charge's EM
interaction with its own EM\ field, and we think that the simplest
relativisic-invariant way\ to provide a seamless balance in all regimes is
to take a balancing adjacent field which satisfies the same Maxwell's
equations as the EM field \ and enters the "charges-fields" Lagrangian
anti-symmetrically with the EM\ field. The system (\ref{fmunu})-(\ref{KG0})
\ determines the dynamics of the fields $\psi ,F^{\mu \nu },F^{\mathrm{ad}%
\mu \nu }$. Since the $\psi $ satisfies the KG\ equation, \ the continuity
equation is fulfilled for the current $J^{\nu }$, and therefore the Lorentz
gauge on the fields $F^{\mu \nu },F^{\mathrm{ad}\mu \nu }$ is preserved in
the dynamics. Therefore the Maxwell equations are effectively reduced to
wave equations for the potentials $A^{\nu },A^{\mathrm{ad}\nu }$ \ coupled
with the KG\ equation through the covariant derivatives. \ The Maxwell
equations have many solutions, and we want to show that a certain class of
solutions can be singled out so that Dirac's analysis can be applied to
them. We single out a class of solutions of Maxwell equations for the EM
field and the adjacent fields by determining them in terms of Green
functions. The EM potentials are determined in terms of the the current $%
J^{\nu }$ \ by the formula 
\begin{equation}
A^{\nu }=\vartheta _{0}\mathcal{G}_{\mathrm{ret}}J^{\nu }+\left( 1-\vartheta
_{0}\right) \mathcal{G}_{\mathrm{adv}}J^{\nu },  \label{aj}
\end{equation}%
where $\mathcal{G}_{\mathrm{ret}}$, $\mathcal{G}_{\mathrm{adv}}$ \ are the
solution operators of Maxwell equations in terms of the retarded and
advanced potentials respectively, the parameter $\vartheta _{0}=1$ for
causal solutions and $\vartheta _{0}=1/2$ for time-symmetric solutions as in
Wheeler-Feynman theory. \ One of our goals is to show that the radiative
response of a point charge depends on the extended regularization of the
point charge. Therefore, we introduce a family of\ adjacent fields which
depend on a parameter, namely the adjacent potentials are defined by the
formula 
\begin{equation}
A^{\mathrm{ad}\nu }=\vartheta _{1}\mathcal{G}_{\mathrm{ret}}J^{\nu }+\left(
1-\vartheta _{1}\right) \mathcal{G}_{\mathrm{adv}}J^{\nu },  \label{aadj}
\end{equation}%
where $\vartheta _{1}$ is a real parameter.

Now we briefly discuss the structure of the system (\ref{fmunu})-(\ref{KG0}%
). Maxwell equations (\ref{fmunu}), (\ref{fadmunu}) are coupled with the KG\
equation (\ref{KG0}) through the covariant derivatives (\ref{comsemh}) which
involve the difference $A^{\mu }-A^{\mathrm{ad}\mu }$ of the potentials $%
A^{\mu },A^{\mathrm{ad}\mu }$. The KG equation (\ref{KG0}) with the
logarithmic nonlinearity $G^{\prime }$ and a given field $A^{\mu }-A^{%
\mathrm{ad}\mu }$ is well-posed, \cite{CazenaveHaraux80}. If $A^{\mathrm{ad}%
\mu }$ is given, the subsystem (\ref{fmunu}), (\ref{KG0}) is also
well-posed, \cite{LongS08}. Note that if we introduce the adjacent field as
a variable, the well-posedness is even better. The difference $A^{\mu }-A^{%
\mathrm{ad}\mu }$ is a solution of the homogenious Maxwell equation with
given Cauchy data and is uniquely defined by the data. Substituting $A^{\mu
}-A^{\mathrm{ad}\mu }$ in the KG equation (\ref{KG0}),\ we find $\psi $ and
then determine $J^{\nu }$ and finally $A^{\mu },A^{\mathrm{ad}\mu }$.
Therefore, the Cauchy data for $\psi ,A^{\mu },A^{\mathrm{ad}\mu }$ uniquely
determine the solution to (\ref{fmunu})-(\ref{KG0}), and the problem of
determination of solutions to (\ref{fmunu})-(\ref{KG0}) is well-posed. Now
we turn to the assumptions (\ref{aj})-(\ref{aadj}). They can be looked at
from two perspectives. The first approach is to simply consider formulas (%
\ref{aj})-(\ref{aadj}) as a way to describe a special subset of solutions of
(\ref{fmunu})-(\ref{KG0}). This subset of solutions defines a corresponding
set $\mathcal{S}$ of Cauchy data for $\psi ,A^{\mu },A^{\mathrm{ad}\mu }$.
If we take Cauchy data from $\mathcal{S}$, the solution of the system (\ref%
{fmunu})-(\ref{KG0}) will automatically satisfy formulas (\ref{aj})-(\ref%
{aadj}), but the process of solving the equations is local in time and does
not involve the formulas (\ref{aj})-(\ref{aadj}). Still the restriction to
the set $\mathcal{S}$ may lead to subtler problems with causality, as can be
examplified by the LAD equation, where the restriction to solutions without
the run-off leads to the pre-acceleration. Importantly, if the potentials $%
A^{\mu },A^{\mathrm{ad}\mu }$ are defined by (\ref{aj})-(\ref{aadj}), the
potential $A^{\mu }-A^{\mathrm{ad}\mu },$ which enters (\ref{KG0}),\ is
exactly proportional to the potential of Dirac's radiation field $F_{\mathrm{%
rad}}^{\mu \nu }=F_{\mathrm{ret}}^{\mu \nu }-F_{\mathrm{adv}}^{\mu \nu }$
which plays a significant role in Dirac's analysis \cite{Dirac CE}; this
shows the relevance of the class of solutions given by (\ref{aj})-(\ref{aadj}%
) \ for the treatment of Dirac's method in our framework. The second
approach to (\ref{fmunu})-(\ref{aadj}) is elimination of $A^{\mu },A^{%
\mathrm{ad}\mu }$ as independent variables. We can use (\ref{aj})-(\ref{aadj}%
) to express $A^{\mu },A^{\mathrm{ad}\mu }$ in terms of $\psi $ and obtain
an equation for $\psi $ alone. Now we describe the structure of the set of
solutions. For any Cauchy data $\ \mathbf{\psi }_{0}$ for $\psi $ and any
field $A_{0}^{\mu }-A_{0}^{\mathrm{ad}\mu }$ we find a solution $\psi $ of (%
\ref{KG0}), then determine $J^{\nu }$ in terms of $\psi $ and then find
potentials $A^{\mu },A^{\mathrm{ad}\mu }$ by (\ref{aj})-(\ref{aadj}). The
potentials $A^{\mu },A^{\mathrm{ad}\mu }$ depend on $A_{0}^{\mu },A_{0}^{%
\mathrm{ad}\mu },\mathbf{\psi }_{0}$. If $A^{\mu }-A^{\mathrm{ad}\mu
}=A_{0}^{\mu }-A_{0}^{\mathrm{ad}\mu }$, we obtain a fixed point \ which
determines a solution of (\ref{KG0})-(\ref{aadj}) with the Cauchy data $%
\mathbf{\psi }_{0}$. Note that if $\vartheta _{0}-\vartheta _{1}=0,$ then $%
A^{\nu }-A^{\mathrm{ad}\nu }=0$, and we first find $\psi $ which is
independent of $A^{\nu },A^{\mathrm{ad}\nu }$ and then calculate $A^{\mu
},A^{\mathrm{ad}\mu }$; hence the problem is well-posed. For small $%
\vartheta _{0}-\vartheta _{1}$ we have a small perturbation of this
well-posed problem and may expect that it is well-posed too; perturbative
techniques can be used to find solutions. The information on solutions of (%
\ref{KG0})-(\ref{aadj}) can be used to characterize the set $\mathcal{S}$
introduced in the first approach. We see in particular that $\mathbf{\psi }%
_{0}$ can be prescribed, and Cauchy data for $A^{\mu },A^{\mathrm{ad}\mu }$
should be calculated. The system (\ref{KG0})-(\ref{aadj}) allows to
eliminate the variables $A^{\nu },A^{\mathrm{ad}\nu }$, \ but it is
non-local in time and in the case $\vartheta _{1}\neq 1,\vartheta _{0}\neq 1$
\ explicitly involves advanced potentials \ and meets the same type of
problems with causality as in \cite{Dirac CE}, \cite{Wheeler Feynman 1}, 
\cite{Wheeler Feynman 2}.

We show in this paper that in the point localization limit the trajectory of
the charge satisfies the generalized LAD equation (\ref{vw1}), and this
equation involves a coefficient $\theta $ which depends on the parameters $%
\vartheta _{0},\vartheta _{1}$. Therefore, Dirac's mass renormalization
method can be realized in a rigorous model, and at the same time the
dynamics of a point charge "remembers" the way the point charge was obtained
from a distributed charge. There is an essential difference between EM
fields and adjacent fields, the \emph{adjacent field is not observable }
since it does not affect dynamics of a test charge, see Section \ref{Smany}.
The difference is even bigger after taking the point localization limit.
After taking the limit, a distributed charge turns into a point without any
internal dynamics, and the adjacent field of the\emph{\ }charge, though it
has a non-trivial limit, does not affect dynamics of the charge itself. In a
contrast,\ \ the EM field persists as a field generated by point charges and
is observable through interactions with charges, in particular it affects a
test charge, see Section \ref{Smany}. Now we point to a crucial difference
with Dirac's argument. An important part of his derivation is subtraction of
the half-sum $\frac{1}{2}\left( F_{\mathrm{ret}}^{\mu \nu }+F_{\mathrm{adv}%
}^{\mu \nu }\right) $ of the advanced and retarded fields from the actual
field which acts on the point charge, the subtraction is used as a tool to
remove the singularity of the actual field. This regularizing field as well
as Dirac's non-singular EM radiation field $F_{\mathrm{ret}}^{\mu \nu }-F_{%
\mathrm{adv}}^{\mu \nu }$ of the point charge are \emph{observable} and
involve advanced potentials leading to some problems with causality. The
causal solutions of Maxwell equations for EM fields correspond to the value $%
\vartheta _{0}=1$ for the parameter $\vartheta _{0}$, and in this case the
EM field generated by the point charge is described in terms only of the
retarded potential. In our model, the advanced potentials enter only through
an \emph{unobservable} adjacent field of the point charge and only describe
a specific choice of obtaining a point charge from a distributed charge.
Though the contradiction with causality is not completely eliminated in the
point limit, it is more difficult to observe.  \emph{The only (but
important) memory left of the adjacent field is the value of the coefficient 
}$\vartheta $\emph{\ at the radiation response in the generalized LAD
equation and the contribution to the energy-momentum balance, in particular
the compensation of the infinite Coulomb self-energy of the point charge. }%
The observable EM field of the charge is given only by the retarded EM field%
\emph{\ }$F_{\mathrm{ret}}^{\mu \nu }$, but the singularity of $F_{\mathrm{%
ret}}^{\mu \nu }$ at the location of the charge does not cause problems
since the charge self action before the point limit was compensated by the
adjacent field $F^{\mathrm{ad}\mu \nu }$. Therefore the renormalization of a
point charge via subtracting its infinite EM self-energy combined with the
fulfillment of the LAD equation and the use of its retarded EM\ field as the
field generated by the charge can be completely justified in the framework
of a relativistic field theory if we treat a point charge as a limit of a
properly defined distributed charge. The classical Maxwell-Lorentz EM\
theory of point charges with $\theta =1/2$ therefore is self-consistent if
the point charge is properly introduced. Similarly, theories with the
generalized LAD\ equation which correspond to any fixed value of $\theta $
are also self-consistent, in particular with a negative value of $\theta $
which does not produce runaway solutions of LAD\ equation. Note though that
even if the EM\ theory with point charges is self-consistent, it does not
describe correctly non-classical effects such as discreteness of hydrogen
energies. In a contrast, a theory with distributed charges as in \ \cite{BF5}%
, \cite{BF6}, \cite{BF7} allows to describe such effects.

Note that our derivation of the LAD equation uses only conservation laws for
the involved fields and does not use specific information on the Lagrangian
and corresponding field equations. The derivation in the spirit of Ehrenfest
theorem allows to circumvent technical difficulties which arise if one uses
the ansatz method to study soliton-like solutions of the NKG-Maxwell system
as in \cite{LongS08}. We use two structural properties of the Lagrangian:
(i) the existence of the symmetric energy-momentum tensor (which can be
constructed for relativistic invariant Lagrangians \cite{Barut}, \cite%
{Lanczos VPM}, \cite{LandauLif F}); (ii) the gauge invariance of the
Lagrangian which implies fulfillment of the continuity equation and
conservation of the total charge for a charge distribution. We derive the
relativistic point dynamics from the dynamics of distributed fields in the
localization limit when the energy and charge densities converge to
delta-functions. \emph{The main assumption we impose on the fields is that
the energy of the system "charge-fields" asymptotically concentrates at a
trajectory} $\mathbf{\hat{r}}\left( t\right) $. The concentration can be
described by two microscopic scales: charge size scale $a$ and intermediate
confinement scale $R$; both $a$ and $R$ \ are vanishingly small compared
with the macroscopic scale of order 1. The charge distribution $\psi $ is
essentially confined in a ball of radius $a$ centered at $\mathbf{\hat{r}}%
\left( t\right) $ and asymptotically vanish on a sphere $\partial B_{R}$ of
a larger radius $R,$ where $a/R$ is assumed to be vanishingly small. The
energy $\mathcal{\bar{E}}$ of \ the "charge-fields" system which is confined
in the ball $B_{R}$ \ of radius $R$ \ converges: $\mathcal{\bar{E}}%
\rightarrow \mathcal{\bar{E}}_{\infty }\left( t\right) $ \ in the
localization limit 
\begin{equation}
a\rightarrow 0,\quad R\rightarrow 0,\quad \frac{a}{R}\rightarrow 0.
\label{arlim}
\end{equation}%
We do not assume that in the above limit the quantities $a,R\ $are
arbitrary, they may be subjected to additional restrictions, we only need
that there exists a sequence $a_{n},R_{n}$ which satisfies (\ref{arlim}). We
also assume that on the surface of the sphere $\partial B_{R}$ the EM field $%
F^{\mu \nu }$ \ and the adjacent field $F^{\mathrm{ad}\mu \nu }$ generated
by the charge localized at $\mathbf{\hat{r}}\left( t\right) $ in the regime $%
\frac{a}{R}\rightarrow 0$ can be approximated by solutions of the Maxwell
equations with point sources located at the trajectory $\mathbf{\hat{r}}%
\left( t\right) $. Obviously, the above conditions can be understood as a
qualitative formulation of the assumption that the current densities and the
energy density converge to Dirac's delta-functions, in particular the energy
density converges to the delta-function $\mathcal{\bar{E}}_{\infty }\left(
t\right) \delta \left( \mathbf{x}-\mathbf{\hat{r}}\left( t\right) \right) $.
Such assumptions are natural for point-like behavior of a charge, examples
of their fulfillment in non-trivial regimes are given in \cite{BF9}, \cite%
{BF10}. We do not make any assumptions on the mass of a charge, the
Newtonian mass is derived from the equations. \ An important assumption on
the behavior of the external EM\ field \ is that it varies significantly
only at macroscopic spatial scales, that is at spatial and time scales of
order 1 and not at microscopic scales. We want to stress that this
assumption is relevant and the LAD equation would not describe charge
behavior in our model if the external field strongly varied at microscopic
scales as, for example, Coulomb field does near its singularity. Namely, our
analysis of a distributed charge in Coulomb field as in hydrogen atom, \cite%
{BF5}, \cite{BF6}, \cite{BF7}, shows its non-classical behavior, in
particular discreteness of energy levels. We want to emphasize that the
"non-classical" effects emerge in the classical framework and are not based
on quantum-mechanical considerations.

\textbf{The result of the analysis:}

Under the above assumptions of concentration, the limit energy $\mathcal{%
\bar{E}}_{\infty }\left( t\right) $ \ must satisfy the relativistic formula%
\begin{equation}
\mathcal{\bar{E}}_{\infty }=M_{0}\left( 1-\left\vert \partial _{t}\mathbf{%
\hat{r}}\right\vert ^{2}\right) ^{-1/2}=M_{0}/\left\vert \partial
_{s}t\right\vert  \label{einf}
\end{equation}%
with a constant $M_{0}$ which is interpreted as the rest mass of a charge. \
The 4-trajectory $z\left( s\right) =\left( t\left( s\right) ,\mathbf{\hat{r}}%
\left( t\left( s\right) \right) \right) $ \ must satisfy the generalized
LAD\ equation%
\begin{equation}
M_{0}\dot{v}_{\mu }=f_{\mathrm{ex}\mu }+2\vartheta \left[ \frac{2}{3}q^{2}%
\ddot{v}_{\mu }+\frac{2}{3}q^{2}\left( \dot{v}\dot{v}\right) v_{\mu }\right]
\label{ALDeqt}
\end{equation}%
where 
\begin{equation*}
\vartheta =\vartheta _{0}-\vartheta _{1},
\end{equation*}%
$v=\partial _{s}z=\dot{z},$ $\ $ $f_{\mathrm{ex}\mu }=qv_{\nu }F_{\mathrm{ex}%
\mu }^{\nu }$ is the Lorentz force generated by the external EM\ field and $%
q $ is the total charge of the charge distribution. The parameters $q$ and $%
M_{0}$ for the electron\ take the values $q=e$ and $M_{0}=m$ as in (\ref%
{ALDeq}). \ Note that the equation is completely determined by the
assumptions we made. \ The equation involves the radiative response
coefficient $2\vartheta =2\vartheta _{0}-2\vartheta _{1},$ where $\vartheta
_{0},\vartheta _{1}$\ in (\ref{aj}), (\ref{aadj}) \ describe the class of
solutions of Maxwell's equations given by (\ref{fmunu}), (\ref{fadmunu}).
Looking at the derivation of the point charge dynamics, we can see that all
the principal assumptions made by Dirac, namely the fulfillment of
conservation laws, the possibility to remove the EM field singularity, and
the possibility of mass renormalization \ can be observed. But now they are
not assumptions, but the properties of the model of the charge which are
derived by mathematical analysis of the model.

Now we discuss implications of choosing specific values of $\vartheta _{0}$
and $\vartheta _{1}$. As we already mentioned, the value $\vartheta _{0}=1$
describes the theory where the EM interaction between all the charges of the
system (including itself) is retarded, this case directly agrees with the
principle of causality. The adjacent potential $A^{\mathrm{ad}\nu }$ \
solely describes an internal interaction of an elementary charge such as an
electron with itself and is not observable (see Section \ref{Smany} for a
discussion); it is used only to provide an extended regularization for a
point charge and, and in the localization limit the charge is described by a
point (as Dirac assumes from the very beginning); therefore its internal
dynamics disappears. Therefore, as long as we are interested in point charge
dynamics, causality arguments do not impose restrictions on the choice of
the coefficient $\vartheta _{1}$ which determines the relation between
retarded and advanced adjacent potentials, and the value of $\vartheta _{1}$
is arbitrary. Still there are attractive choices for $\vartheta _{1}$ based
on symmetry. \ A time-reversal symmetric choice $\vartheta _{1}=1/2$ in (\ref%
{aadj}) results in $\vartheta =1/2$, in this case the adjacent field $F^{%
\mathrm{ad}\mu \nu }$ has the same form as Dirac's regularizing field $\frac{%
1}{2}\left( F_{\mathrm{ret}}^{\mu \nu }+F_{\mathrm{adv}}^{\mu \nu }\right) $
and the \emph{generalized LAD\ equation (\ref{ALDeqt}) with }$\vartheta =1/2$%
\emph{\ coincides with the classical LAD\ equation (\ref{ALDeq})}. In
addition to the symmetry, this choice has another, more important advantage.
The local and global energy and momemtum conservation laws are fulfilled for
the system "charge field---EM\ field---adjacent field" (CF-EMF-AF), and
after the localization limit for the system "point charge---EM\
field---adjacent field"(PC-EMF-AF). This is crucial for the cancellation of
the infinite energy of the EM self-interaction. The system "point charge
field---EM\ field"(PC-EMF) is dynamically closed and can be considered as a
subsystem of the system PC-EMF-AF, but the removal of AF can break the
energy balance. Now we look for the case where the influence of the adjacent
field on the energy balance is minimal. Of course, we cannot exclude the
influence of the adjacent field completely if we need the renormalization,
because the renormalization is exactly a modification of the energy balance,
but we can minimize its effects far from the charge. According to Larmor
formula, the radiation power is odd with respect to time inversion,
therefore in the time-symmetric case $\vartheta _{1}=1/2$ the amount of
outcoming from the point radiated energy of the adjacent field equals the
amount of incoming radiated energy, hence the radiated EM energy is balanced
only by the charge radiation reaction; the adjacent field \ can be ignored
in the radiated EM\ energy balance, and from the point of view of radiation
balance the adjacent field is "hidden". \emph{Hence the classical value }$%
\vartheta =1/2$\emph{\ can be characterized as the value for which the
radiated energy balance does not involve the adjacent field.} Therefore, if
we take the localization limit in the case $\vartheta _{1}=1/2$, we obtain
as the result the classical LAD equation with a uniquely defined
renormalized mass, the field of the charge is described by the retarded
potentials, the radiated EM energy balance is fulfilled even if we forget
about the adjacent field, the charge energy is finite if we take into
account the adjacent field, and the adjacent field does not affect Newtonian
dynamics of the point charge. From our point of view, it is still preferable
to be logical and not to treat a quantity which balances the infinite energy
of the charge as nonexistent even if it is almost completely hidden. Now we
discuss the radiated energy balance. Note that the generalized LAD equation
is derived as a necessary condition for the \emph{local }energy-momentum
balance in a contracting vicinity of the point charge, whereas the
determination of the specific value $\vartheta _{1}=1/2$ which leads to the
classical value $\vartheta =1/2$ is based on the \emph{global} condition of
radiative balance. There is a significant difference between the local and
global conditions. The generalized LAD equation describes the dynamics of a
charge in a regular external EM\ field, its derivation is based on the local
energy-momentum conservation laws in a small vicinity of the charge
trajectory. Since the external field acting on the charge is generated by
other charges, the assumption of regularity of the external field means that
the charge remains at a \emph{non-zero distance} from all other charges. The
radiation component of the EM\ field of the charge becomes dominant at a 
\emph{large distance }from the charge where the radiated field interacts
with other charges. Now a few words concerning pre-acceleration. Note that
we derive the LAD equation for macroscopic regular trajectories $\mathbf{%
\hat{r}}\left( t\right) $, and since runoff solutions of the LAD equation
are not macroscopically regular, the trajectories $\mathbf{\hat{r}}\left(
t\right) $ in the case $\vartheta >0$ are not described by arbitrary
solutions of the LAD equation but by special solutions which satisfy Dirac's
condition of finite acceleration at $t\rightarrow \infty $. Such solutions
are known to demonstrate the pre-acceleration at microscopic time scales.
Since the EM field of the charge is causal, one could expect that the
dynamics of the point charge should be causal too, but the pre-acceleration
demonstrates the loss of causality at microscopic scales. It is not
surprising though, since the hidden adjacent field involves the advanced
potential.

In addition to the most attractive choice $\vartheta _{1}=1/2$, we consider
briefly other possibilities.\ If $\vartheta _{1}=1$, we obtain $\vartheta =0$%
, and the adjacent field completely balances the EM field in all regimes,
and there is no EM\ self-interaction as in Wheeler-Feynman theory and no
radiative response for a single charge at all. Note that the observable
radiation of the charge does not vanish, it is described as always by Larmor
formula, and as always energy and momentum conservation laws for the
complete system PC-EMF-AF are fulfilled. If $\vartheta _{1}>1$, we obtain a
negative value $\vartheta <0$ \ of the radiative response coefficient. For
negative values $\vartheta <0$ the generalized LAD\ equation has different
properties compared to the classical LAD equation. The runaway solutions of
the classical LAD equation \ turn into rapidly stabilising solutions, and
the pre-acceleration does not happen. Note that in the case $\vartheta
_{1}\neq 1/2$ the adjacent field contributes to the balance of radiated
energy, and since the adjacent field of a charge does not interact with
other charges, this radiated adjacent energy far from the charge becomes a
"dark energy".

The second natural value for the parameter $\vartheta _{0}$ which describes
the EM field is $\vartheta _{0}=1/2$ as in Wheeler-Feynman theory, \cite%
{Wheeler Feynman 1}, \cite{Wheeler Feynman 2}. If $\vartheta _{1}=1/2$ too,
then $\vartheta =0$ and we obtain a theory which at macroscopic scales
coincides with Wheeler-Feynman theory without self-interaction and without
radiative response for a single charge. This case is the most symmetric with
respect to the time inversion. The description of the radiative response in
this case would require to consider a system of very many interacting
charges as in the universal absorber theory. But if we take $\vartheta
_{1}=1/2-\vartheta ,$ we obtain a modified Wheeler-Feynman theory with a
non-zero self-interaction in accelerating regimes and a non-zero radiative
response for a single charge which does not rely on the existence of the
universal absorber. The value of $\vartheta $ can be positive or negative.
If $\vartheta _{1}=1$, that is the adjacent field is retarded, then $%
\vartheta =-1/2\ \ $and the pre-acceleration does not happen.

The above discussion shows that the radiation response of a point charge
depends on the way the point charge is obtained from distributed charges.
Though the value $\vartheta _{1}=1/2$, which leads to the classical value $%
\vartheta =1/2$ of the radiative response coefficient, is the most
attractive choice, there is no purely mathematical argument which excludes
all other choices. A purely field-theoretic analysis without additional
assumptions does not uniquely define the radiative response coefficient,\
and its determination should rely on experimental data. The experimental
methods for verification of the LAD equation described in \cite[Section 9.3]%
{Spohn}, could be used for the determination of the value of $\vartheta $. \
At the same time ,our analysis shows that classical electrodynamics is
self-consistent if the concept of an elementary point charge is properly
introduced, which agrees with conclusions made in \cite{Rohrlich} on
different grounds.

\textbf{Conclusions:}

(i) All assumptions of Dirac's analysis (fulfillment of local conservation
laws, EM field singularity removal, mass renormalization) can be
consistently substantiated by a properly introduced model of a point charge.

(ii) Introduction of a point charge as a limit of distributed charges in the
framework of a relativistic Lagrangian field theory allows to consistently
derive the relativistic version of Newton's equations for the point charge,
Einstein's formula and generalized LAD\ equation. The derivation of the
generalized LAD equation is based only on local conservation laws and does
not involve assumptions of global nature.

(iii) The mass renormalization is explicitly demonstrated by the
cancellation of two infinite energies in the dynamical regimes: the energy
of the EM field and the energy of the adjacent field with uniquely
determined renormalized energy and mass.

(iv) The magnitude and sign of the radiative response of a charge is
affected by the composition of Poincar\'{e} cohesive forces (through the
dependence on $\vartheta $). The nonlinearity does not affect directly the
radiative response, though it affects the rest mass $m$ of the charge. The
rest mass does not depend on $\vartheta $.

(v) The observable radiation of a charge which may affect a test charge is
described by the EM field of the point charge and can be calculated by
classical formulas.

(vi) The classical field-theoretic approach allows to consistently implement
the mass renormalization and derive the radiation response described by the
generalized LAD equation, but the value of the radiative response
coefficient $\vartheta $ and in particular its sign is not determined by
local energy-momentum conservation laws.

(vii) The classical value $\vartheta =1/2$ is selected by the global
requirement that the adjacent field of the charge does not contribute to the
balance of radiated energy.

In this paper we consider the simplest model of a charge described by a
scalar complex field $\psi $. A particle with spin 1/2 can be quite
similarly described by a spinor field. Since our analysis does not use
specific structure of field equations and the energy-momentum tensor and
only uses its symmetry and conservation laws, our approach is applicable in
this case too.

\subsection{Lagrangian formalism and field equations}

We use the following notation for the time-space 4-vector in its
contravariant $x^{\mu }$ and covariant $x_{\mu }$ forms:%
\begin{equation}
x=x^{\mu }=\left( x^{0},x^{1},x^{2},x^{3}\right) =\left( \mathrm{c}t,\mathbf{%
x}\right) ,\ \ x_{\mu }=g_{\mu \nu }x^{\nu }=\left( \mathrm{c}t,-\mathbf{x}%
\right) ,  \label{fre4}
\end{equation}%
\begin{equation}
\partial _{\mu }=\partial /\partial x^{\mu }=\left( \mathrm{c}^{-1}\partial
_{t},\nabla \right) ,\ \partial ^{\mu }=\partial /\partial x_{\mu }=\left( 
\mathrm{c}^{-1}\partial _{t},-\nabla \right) ,  \label{fre6}
\end{equation}%
we set the speed of light $\mathrm{c}=1$, and we use the common convention
on the summation over the same indices. We use notation for the space
3-vector $\ x^{i}=\left( x^{1},x^{2},x^{3}\right) =\mathbf{x},\ i=1,2,3,$
and we \ emphasize notationally by the Latin superscript its difference from
4-vector $x^{\mu }$ with the Greek superscript. The metric tensor $g_{\mu
\nu }$ is defined by%
\begin{equation}
\left\{ g_{\mu \nu }\right\} =\left\{ g^{\mu \nu }\right\} =diag\left[
1,-1,-1,-1\right] .  \label{fre7}
\end{equation}

We consider the "charge-fields" system which describes the interaction of a
single distributed charge with three fields: its own EM field, the adjacent
field generating Poincar\'{e} cohesive forces and the external EM field. The
internal adjacent field of the charge is introduced in the model to balance
the EM\ self-interaction of the charge and provide its stability. The charge
dynamics is described by the complex scalar charge distribution $\psi \left(
x\right) $, the EM field by its 4-potential $A^{\mu }\left( x\right) $, and
the adjacent field\ by the 4-potential $A^{\mathrm{ad}\mu }\left( x\right) $%
; the system also involves a prescribed external field with the 4-potential $%
A_{\mathrm{ex}}^{\mu }$. The adjacent field is solely responsible for the
balancing internal interaction of the charge with itself and does not affect
other charges (see Section \ref{Smany} for a discussion of the interaction
structure in the case of many charges).

The system "charge-fields" is described by the Lagrangian

\begin{equation}
\mathcal{L}=L_{\mathrm{KG}}-L_{\mathrm{em}}\left( A\right) +L_{\mathrm{em}%
}\left( A^{\mathrm{ad}}\right) .  \label{flagr6as}
\end{equation}%
where we for brevity set $A=A^{\mu },A^{\mathrm{ad}}=A^{\mathrm{ad}\mu }$.
Here $L_{\mathrm{KG}}$ is the Lagrangian of a nonlinear Klein-Gordon
equation: 
\begin{equation}
L_{\mathrm{KG}}=\frac{\hbar ^{2}}{2m}\left\{ \psi _{;\mu }\psi ^{;\mu }-%
\frac{m^{2}}{\hbar ^{2}}\psi ^{\ast }\psi -G\left( \psi ^{\ast }\psi \right)
\right\} ,  \label{paf1h}
\end{equation}%
where the star denotes the complex conjugation, $m$ is a mass parameter, it
is related to the mass of the charge, see \cite{BF8}, \cite{BF9} and Section %
\ref{Srest}. It involves covariant derivatives 
\begin{equation}
\psi ^{;\mu }=\tilde{\partial}^{\mu }\psi =\partial ^{\mu }\psi +\frac{%
\mathrm{i}q}{\hbar }\tilde{A}^{\mu }\psi ,  \label{comsemh}
\end{equation}%
\begin{equation}
\tilde{\partial}^{\mu }=\partial ^{\mu }+\frac{\mathrm{i}q}{\hbar }\tilde{A}%
^{\mu },\ \ \partial ^{\mu }=\frac{\partial }{\partial x_{\mu }}=\left(
\partial _{t},-\nabla \right) ,\ \ \partial _{\mu }=\frac{\partial }{%
\partial x^{\mu }}=\left( \partial _{t},\nabla \right) ,
\end{equation}%
where $q$ is the charge parameter, and the total field potential $\tilde{A}%
^{\mu }$ affecting the charge is defined by 
\begin{equation}
\tilde{A}^{\mu }=A_{\mathrm{ex}}^{\mu }+A^{\mu }-A^{\mathrm{ad}\mu }
\label{aaex}
\end{equation}%
where $A_{\mathrm{ex}}^{\mu }$ is the potential of the external EM field.
The Lagrangians $L_{\mathrm{em}}\left( A\right) $, $L_{\mathrm{em}}\left( A^{%
\mathrm{ad}}\right) $ for the EM\ field and the adjacent field are
Maxwellian: 
\begin{equation}
L_{\mathrm{em}}\left( A\right) =-\frac{1}{16\pi }F_{\mu \nu }F^{\mu \nu
},\quad L_{\mathrm{em}}\left( A^{\mathrm{ad}}\right) =-\frac{1}{16\pi }%
F_{\mu \nu }^{\mathrm{ad}}F^{\mathrm{ad}\mu \nu }  \label{Lem}
\end{equation}%
with%
\begin{equation}
\ F^{\mu \nu }=\partial ^{\mu }A^{\nu }-\partial ^{\nu }A^{\mu },\ \ \ F^{%
\mathrm{ad}\mu \nu }=\partial ^{\mu }A^{\mathrm{ad}\nu }-\partial ^{\nu }A^{%
\mathrm{ad}\mu }.  \label{Fmunug}
\end{equation}%
Obviously, the Lagrangian $\mathcal{L}$ is relativistic invariant. The field
equations which describe dynamics of the charge-fields system are obtained
as Euler-Lagrange equations for the Lagrangian $\mathcal{L}$ and involve a
nonlinear Klein-Gordon (KG) equation for the charge distribution $\psi $ and
Maxwell equations for the potentials. The KG equation has the form (\ref{KG0}%
), namely 
\begin{equation}
\tilde{\partial}^{\mu }\tilde{\partial}^{\mu }\psi +\frac{m^{2}}{\hbar ^{2}}%
\psi +G^{\prime }\left( \psi ^{\ast }\psi \right) \psi =0  \label{KGone}
\end{equation}%
where $G^{\prime }\left( s\right) =dG/ds$. The equations for the fields $\
F^{\mu \nu },F^{\mathrm{ad}\mu \nu }$ take the form \ of Maxwell equations \
(\ref{fmunu}), (\ref{fadmunu}) \ where the source current is defined by%
\begin{equation}
J^{\nu }=-\frac{\partial L_{\mathrm{KG}}}{\partial A_{\nu }}=-\mathrm{i}%
\frac{q}{\hbar }\left( \frac{\partial L_{\mathrm{KG}}}{\partial \psi _{;\nu }%
}\psi -\frac{\partial L_{\mathrm{KG}}}{\partial \psi _{;\nu }^{\ast }}\psi
^{\ast }\right) .  \label{Jnu}
\end{equation}%
The nonlinearity $G$ is required to provide the localization of the charge
and involves dependence on the size parameter $a$: $G\left( b\right)
=a^{-5}G\left( a^{3}b\right) $, the parameter $a$ determines the scale of
localization. Note that a particular form of the nonlinearity is not
important in our analysis here. A typical example of the nonlinear term $%
G^{\prime }$ in the KG\ equation is given by the logarithmic expression,%
\begin{equation}
G^{\prime }\left( b\right) =-a^{-2}[\ln \left( a^{3}b\right) +\ln \pi
^{3/2}+3],\quad b=\left\vert \psi \right\vert ^{2}\geq 0,  \label{paf30}
\end{equation}%
where $a$ is the \emph{charge size parameter}. \ For this choice of the
nonlinearity the free resting charge 
\begin{equation}
\psi =\mathrm{e}^{-\mathrm{i}tm/\hbar }\left\vert \psi \right\vert
\label{phar}
\end{equation}%
has a Gaussian shape, namely 
\begin{equation}
\left\vert \psi \right\vert =\pi ^{-3/4}a^{-3/2}\mathrm{e}^{-\left\vert 
\mathbf{x}\right\vert ^{2}a^{-2}/2},  \label{modp}
\end{equation}%
and the parameter $a$ \ can be interpreted as the size of the free charge.
The nonlinear KG\ equation with the logarithmic nonlinearity has\ good
dynamical properties, \cite{CazenaveHaraux80}. The KG Lagrangian is gauge
invariant, \ namely for any real $\gamma $ \ 
\begin{equation}
L_{\mathrm{KG}}\left( \mathrm{e}^{\mathrm{i}\gamma }\psi ,\mathrm{e}^{%
\mathrm{i}\gamma }\psi _{;\mu },\mathrm{e}^{-\mathrm{i}\gamma }\psi ^{\ast },%
\mathrm{e}^{-\mathrm{i}\gamma }\psi _{;\mu }^{\ast }\right) =L_{\mathrm{KG}%
}\left( \psi ,\psi _{;\mu },\psi ^{\ast },\psi _{;\mu }^{\ast }\right) .
\label{eisinv}
\end{equation}%
Therefore the current $J^{\nu }=\left( \rho ,\mathbf{J}\right) $ satisfies
the charge conservation/continuity equation%
\begin{equation}
\partial _{\nu }J^{\nu }=0\text{ }  \label{cont1}
\end{equation}%
which can be written in the form 
\begin{equation}
\partial _{t}\rho +\nabla \cdot \mathbf{J}=0.  \label{paf6a}
\end{equation}%
The components of the 4-current defined by (\ref{Jnu}) for the KG equation
have the form 
\begin{equation}
\rho =-\frac{\hbar q}{m}\left\vert \psi \right\vert ^{2}\func{Im}(\tilde{%
\partial}_{t}\psi /\psi ),\quad \mathbf{J}=\frac{\hbar q}{m}\left\vert \psi
\right\vert ^{2}\func{Im}(\tilde{\nabla}\psi /\psi ).  \label{paf6}
\end{equation}%
Using the fulfillment of the continuity equation, we impose Lorentz gauge on
the solutions of Maxwell equations. According to the continuity equation the
total charge $\int_{\mathbb{R}^{3}}\rho d\mathbf{x}$ is preserved. The KG\
equation in the absence of external fields has time-harmonic solutions of
the form (\ref{phar}) where $\left\vert \psi \right\vert $ satisfies the
normalization condition 
\begin{equation*}
\int_{\mathbb{R}^{3}}\left\vert \psi \right\vert ^{2}d\mathbf{x}=1.
\end{equation*}%
An example of $\left\vert \psi \right\vert $ which satisfies the
normaization condition is given by (\ref{modp}). According to (\ref{paf6}), $%
\rho =q\left\vert \psi \right\vert ^{2}$ for such harmonic solutions, and
the normalization condition implies that the potential of the static EM\
field generated by a resting charge approaches the Coulomb potential $\frac{q%
}{\left\vert \mathbf{x}\right\vert }$ as $\left\vert \mathbf{x}\right\vert
\rightarrow \infty $ or $a\rightarrow 0$, i.e. the charge parameter $q$ can
be interpreted as the total charge of the particle.

Now we make a remark on the choice of the Lagrangian. The simplest
relativistic invariant field equation which involves a mass parameter is the
Klein-Gordon equation; to obtain a point-like distribution of the charge we
introduce the nonlinearity. To obtain the EM\ field described by Maxwell's
equations we introduce corresponding Lagrangian. The coupling of the charge
and the EM field through covariant derivatives is the minimal coupling as in 
\cite{Appel Kiessling}, \cite{Nodvik}. The simplest way to balance the EM
field by another field is to introduce the adjacent field into the
Lagrangian anti-symmetrically with the EM field. Therefore we think that the
Lagrangian (\ref{flagr6as}) describes the most natural relativistic
covariant model of a distributed charge interacting with the EM field in a
balanced way which provides stability of the charge and its asymptotic
localization, but still allows non-trivial EM self-action.

\subsection{Energy-momentum tensor and conservation laws \label{ChTensor}}

The symmetric energy-momentum \ tensor (EnMT) $\Theta ^{\mu \nu }$ for the
EM field is given by the formula \cite{Jackson}, \cite{Barut}: 
\begin{equation}
\Theta ^{\mu \nu }\left( F\right) =\frac{1}{4\pi }\left( g^{\mu \alpha
}F_{\alpha \beta }F^{\beta \nu }+\frac{1}{4}g^{\mu \nu }F_{\alpha \beta
}F^{\alpha \beta }\right) ,\   \label{maxw11}
\end{equation}%
and the same formula defines $\Theta ^{\mu \nu }\left( F^{\mathrm{ad}%
}\right) $. The symmetric EnMT \ for the KG equation has the form 
\begin{eqnarray}
T^{\mu \nu } &=&\frac{\partial L_{\mathrm{KG}}}{\partial \psi _{;\mu }}\psi
^{;\nu }+\frac{\partial L_{\mathrm{KG}}}{\partial \psi _{;\mu }^{\ast }}\psi
^{;\nu \ast }-g^{\mu \nu }L_{\mathrm{KG}}  \label{TKG} \\
&=&\frac{\hbar ^{2}}{2m}\left[ \psi ^{;\mu \ast }\psi ^{;\nu }+\psi ^{;\mu
}\psi ^{;\nu \ast }\right] -g^{\mu \nu }L_{\mathrm{KG}}.  \notag
\end{eqnarray}%
The symmetric energy-momentum tensor for the charge-fields system (\ref%
{flagr6as}) has the form 
\begin{equation}
\mathcal{T}^{\mu \nu }=T^{\mu \nu }+\Theta ^{\mu \nu }\left( F\right)
-\Theta ^{\mu \nu }\left( F^{\mathrm{ad}}\right) ,  \label{Tmunutot}
\end{equation}%
where $F,F^{\mathrm{ad}}$ are defined by (\ref{Fmunug}). The crucial
property of the EnMT is its symmetery 
\begin{equation}
\mathcal{T}^{\mu \nu }=\mathcal{T}^{\nu \mu }.  \label{symp}
\end{equation}

The \emph{energy and momentum conservation laws }\ for the system (\ref%
{KGone})-(\ref{fadmunu}) have the following form: \emph{\ }%
\begin{equation}
\partial _{\mu }\mathcal{T}^{\mu \nu }=f^{\nu }.  \label{paf14}
\end{equation}%
Here $f^{\nu }=-\partial \mathcal{L}/\partial x_{\nu }$ \ is the \emph{%
Lorentz force density, }it is expressed in terms of the external EM field $%
F_{\mathrm{ex}}^{\nu \mu }$ acting on the 4-current $J^{\mu }$ as follows: 
\begin{equation}
f^{\nu }=J_{\mu }F_{\mathrm{ex}}^{\nu \mu }=\left( f^{0},\mathbf{f}\right) .
\label{fre15}
\end{equation}%
Fulfillment of (\ref{Tmunutot}) is verified in a standard way\ \cite{Barut}, 
\cite{Lanczos VPM}, \cite{LandauLif F}.

The symmetric EnMT\ involves the energy and momentum components, namely 
\begin{equation*}
u=\mathcal{T}^{00},\quad \mathcal{P}=p^{i}=\mathcal{T}^{0i},\quad \ i=1,2,3.
\end{equation*}%
The EnMT conservation law (\ref{paf14}) with $\nu =0$ produces the energy
conservation law%
\begin{equation}
\partial _{t}u+\partial _{j}\mathcal{T}^{j0}=f^{0}\ ,\text{ }
\label{emten6g}
\end{equation}%
which can be rewritten using (\ref{symp}): 
\begin{equation}
\partial _{t}u+\partial _{j}p^{j}=f^{0}\ \text{ }  \label{equen2g}
\end{equation}%
with%
\begin{equation}
f^{0}=J_{\mu }F_{\mathrm{ex}}^{0\mu }.  \label{f0jf}
\end{equation}%
The momentum equation takes the form%
\begin{equation}
\partial _{t}p^{j}+\partial _{i}\mathcal{T}^{ij}=f^{j},  \label{momeq1g}
\end{equation}%
or%
\begin{equation}
\partial _{t}\mathcal{P}+\partial _{i}\mathcal{T}^{ij}=\mathbf{f},
\label{mom1g}
\end{equation}%
where the Lorentz force density is given by the formula 
\begin{equation}
\mathbf{f}=f^{j}=J_{\mu }F^{j\mu }\left( A_{\mathrm{ex}}\right) =J_{0}F_{%
\mathrm{ex}}^{j0}+J_{i}F_{\mathrm{ex}}^{ji}.  \label{florg}
\end{equation}

\section{Trajectory of concentration}

We understand a point charge as a localization limit of a distributed
charge. The localization means roughly speaking that the charge and energy
densities converge to Dirac's delta-functions. The location of this
delta-function determines the trajectory $\mathbf{\hat{r}}\left( t\right) $
\ of the charge. To derive the law of motion of the charge which should
determine the trajectory, we assume that the energy of the charge
concentrates in a small vicinity of the trajectory,\ and the radius of the
vicinity tends to zero in the localization limit (\ref{arlim}). Our analysis
shows that in a very general situation the fulfillment of energy-momentum
conservation laws and the continuity equation uniquely determines the law of
motion, namely the LAD equation; it also implies Einstein's equivalence of
the concentrated energy and the mass of the charge.

We will show that the assumption that the energy of the charge-fields system
is concentrated in a vanishing neighborhood of a trajectory implies a
differential equation for the trajectory. We introduce the \ value $%
F_{\infty }^{\mu \nu }\left( t\right) $ of the external field $\mathbf{\ }%
F^{\mu \nu }\left( t,\mathbf{x}\right) $ \ restricted to the trajectory $%
\mathbf{\hat{r}}\left( t\right) $:

\begin{equation}
F_{\infty }^{\mu \nu }\left( t\right) =F_{\mathrm{ex}}^{\mu \nu }\left( t,%
\mathbf{\hat{r}}\left( t\right) \right) .  \label{Fmninf}
\end{equation}
\ The total EnMT $\mathcal{T}^{\mu \nu }$ is given by formula (\ref{Tmunutot}%
), which we rewrite in the form%
\begin{equation}
\mathcal{T}^{\mu \nu }=T^{\mu \nu }+\tilde{\Theta},\quad \tilde{\Theta}%
=\Theta ^{\mu \nu }\left( F\right) -\Theta ^{\mu \nu }\left( F^{\mathrm{ad}%
}\right) .  \label{Tone}
\end{equation}%
In particular, the energy density 
\begin{equation}
u=\mathcal{T}^{00}=T^{00}+\Theta ^{00}\left( F\right) -\Theta ^{00}\left( F^{%
\mathrm{ad}}\right) .  \label{uden}
\end{equation}%
\ 

Now we formulate our assumptions (complete mathematical details in a similar
situation can be found in \cite{BF9}, \cite{BF10}). \ We introduce a small
ball 
\begin{equation*}
\Omega =\Omega _{R}=\Omega \left( \mathbf{\hat{r}},R\right) =\left\{ \mathbf{%
x:}\quad \left\vert \mathbf{x-\hat{r}}\left( t\right) \right\vert \leq
R\right\}
\end{equation*}%
centered at a point of the trajectory, below we take $R\rightarrow 0$. We
assume that \ the energy of the charge-fields system \ concentrates at the
trajectory $\mathbf{\hat{r}}\left( t\right) $ on a time interval\ $%
T_{-}<t<T_{+}$. The field variables are split into two groups: a strongly
localized charge distribution $\psi $ \ and EM field potentials $A^{\nu },A^{%
\mathrm{ad}\nu }$. There is a difference in the localization of the charge
distributions and the EM fields. An example of a strongly localized charge
is given by the Gaussian form factor 
\begin{equation}
\left\vert \psi \right\vert ^{2}=\frac{1}{a^{3}\pi ^{3/2}}\exp \left( -\frac{%
1}{a^{2}}\left\vert \mathbf{x}-\mathbf{\hat{r}}\right\vert ^{2}\right)
\label{Gaus}
\end{equation}%
which is strongly localized in $\Omega =\Omega _{R};$ in particular \ the
values of the charge on the sphere 
\begin{equation*}
\partial \Omega =\left\{ \mathbf{x}:\left\vert \mathbf{x}-\mathbf{\hat{r}}%
\right\vert =R\right\}
\end{equation*}%
tend to zero very fast \ if $a/R\rightarrow 0$. The EM fields are not
localized, for instance \ the Coulomb field potential behaves as $\frac{q}{%
\left\vert \mathbf{x}-\mathbf{\hat{r}}\right\vert }$ and its values on $%
\partial \Omega $\ do not vanish even if $a=0$ when the source is completely
localized as a delta-function.

We concider the concentrated in $\Omega $ \ energy%
\begin{equation}
\mathcal{\bar{E}}\left( t\right) =\int_{\Omega }\mathcal{T}^{00}\mathrm{d}%
\mathbf{x}=\int_{\Omega }u\mathrm{d}\mathbf{x},  \label{concen}
\end{equation}%
and the concentrated in $\Omega $ charge%
\begin{equation}
\bar{\rho}\left( t\right) =\int_{\Omega }\rho \mathrm{d}\mathbf{x}.
\label{rhobar}
\end{equation}%
Our main assumption is that the concentrated energy and charge converge \ in
the localization limit (\ref{arlim}): 
\begin{equation}
\mathcal{\bar{E}}\left( t\right) \rightarrow \mathcal{\bar{E}}_{\infty
}\left( t\right) \neq 0,  \label{T00lim}
\end{equation}%
\begin{equation}
\bar{\rho}\left( t\right) \rightarrow \bar{\rho}_{\infty }\left( t\right) .
\label{rhobarlim}
\end{equation}%
It is worth noticing that since $F^{\mu \nu }$ and $F^{\mathrm{ad}\mu \nu }$
are solutions of Maxwell equations with the same source, the difference $%
F^{\mu \nu }-F^{\mathrm{ad}\mu \nu }$ does not involve a singularity;
therefore condition (\ref{T00lim}) mainly concerns behavior of components of
the KG\ tensor $T^{\mu \nu }$.

In addition to the above condition on convergence, we assume that certain
integrals\ asymptotically vanish. There are two types of vanishing
integrals. The first type includes volume integrals over the ball $\Omega $
\ or surface integrals over its boundary $\partial \Omega $ which involve
vanishing factors. In particular, \ since $\left( x^{i}-\hat{r}^{i}\right)
=O\left( R\right) ,$ we assume that 
\begin{equation}
\int_{\Omega }\left( x^{i}-\hat{r}^{i}\right) \rho \mathrm{d}\mathbf{x}%
=o\left( 1\right) ,\quad \int_{\partial \Omega }\left( \mathbf{x}-\mathbf{%
\hat{r}}\right) \mathbf{\hat{v}}\cdot \mathbf{\bar{n}}u\,\mathrm{d}\sigma
=o\left( 1\right)  \label{Tbou}
\end{equation}%
where $\mathbf{\bar{n}}$ is the external normal to the sphere $\partial
\Omega $, $\mathbf{\hat{v}=\partial }_{t}\mathbf{\hat{r}}$ is the
trajectory-based velocity (we denote by $O\left( R^{p}\right) $ such a
quantity that $O\left( R^{p}\right) /R^{p}$ remains bounded in the
localization limit; the notation $o\left( R^{p}\right) $ means that $o\left(
R^{p}\right) /R^{p}\rightarrow 0$). The second type of vanishing integrals
are surface integrals which are negligible because of the strong
localization of $\psi $ as in (\ref{Gaus}). Since the \ radius of
localization of $\psi $ is of order $a$ and $R>>a$ , we assume that surface
integrals over $\partial \Omega $ that involve $\psi $ or its derivatives as
factors asymptotically vanish. For example, for $\rho $ and $\mathbf{J}$
given by (\ref{paf6}), we assume that 
\begin{equation}
\int_{\partial \Omega }\left( \mathbf{\hat{v}}\cdot \mathbf{\bar{n}}\rho -%
\mathbf{\bar{n}}\cdot \mathbf{J}\,\right) \mathrm{d}\sigma =o\left( 1\right)
.  \label{To1}
\end{equation}%
Several surface integrals, which are similar to the above expressions, are
also assumed to asymptotically vanish; they can be easily extracted from the
calculations made in Section \ref{Sconv}. Note that this kind of assumptions
is natural for a point-like behavior of a charge distribution, they are
obviously fulfilled for delta-functions as in (\ref{msource}). Detailed
mathematical formulations made in similar situations and the verification of
fulfillment of this kind of conditions for non-trivial examples can be found
in \cite{BF9}, \cite{BF10}.

The unlocalized quantities which involve only the EM fields $A^{\nu }$ or $%
A^{\mathrm{ad}\nu }$ can make a non-vanishing contribution to the surface
integrals, namely we assume that 
\begin{equation}
\int_{\partial \Omega }\left( \hat{v}^{i}\bar{n}^{i}\mathcal{T}^{00}-\bar{n}%
^{i}p^{i}\right) \,\mathrm{d}\sigma =W_{\mathrm{rad}}+o\left( 1\right) ,
\label{bound3r}
\end{equation}%
\begin{equation}
\int_{\partial \Omega }\left( \hat{v}^{i}\bar{n}^{i}p^{j}-\bar{n}^{i}%
\mathcal{T}^{ij}\right) \,\mathrm{d}\sigma =f_{\mathrm{rad}}^{j}+o\left(
1\right) .  \label{bound6r}
\end{equation}

We make the following assumption of the localization of the charge at
microscopic scales. On the sphere $\left\vert \mathbf{x-\hat{r}}\left(
t\right) \right\vert =R$, \ since $\frac{a}{R}\rightarrow 0$, \ the field
quantities $F^{\mu \nu },F^{\mathrm{ad}\mu \nu }$ obtained as solutions of
Maxwell equations (\ref{flagr8d}), (\ref{flagr8dd}) with currents $J^{\nu
}=\left( \rho ,\mathbf{J}\right) $ can be asymptotically replaced by the
solutions of Maxwell equations with the currents $J_{\infty }^{\nu }=\left(
\rho _{\infty },\mathbf{J}_{\infty }\right) $ \ for the point charges: 
\begin{equation}
\rho _{\infty }\left( t,\mathbf{x}\right) =\bar{\rho}_{\infty }\delta \left(
t,\mathbf{x}-\mathbf{\hat{r}}(t)\right) ,\quad \mathbf{J}_{\infty }\left( t,%
\mathbf{x}\right) =\mathbf{\hat{v}}\bar{\rho}_{\infty }\delta \left( t,%
\mathbf{x}-\mathbf{\hat{r}}(t)\right) .  \label{msource}
\end{equation}%
Namely, we assume convergence on $\partial \Omega =\left\{ \left\vert 
\mathbf{x-\hat{r}}\left( t\right) \right\vert =R\right\} $ \ in the
localization limit (\ref{arlim}) of the advanced and retarded potentials: 
\begin{equation}
\left( \mathcal{G}_{\mathrm{ret}}J^{\nu }\right) _{\mathbf{x}\in \partial
\Omega }\rightarrow \left( \mathcal{G}_{\mathrm{ret}}J_{\infty }^{\nu
}\right) _{\mathbf{x}\in \partial \Omega },\quad \left( \mathcal{G}_{\mathrm{%
adv}}J^{\nu }\right) _{\mathbf{x}\in \partial \Omega }\rightarrow \left( 
\mathcal{G}_{\mathrm{adv}}J_{\infty }^{\nu }\right) _{\mathbf{x}\in \partial
\Omega }.  \label{gretc}
\end{equation}%
We will use the above convergence to evaluate radiative terms $W_{\mathrm{rad%
}}$ and $f_{\mathrm{rad}}^{j}$ in (\ref{bound3r}), (\ref{bound6r}).

\section{Derivation of Newton's law for the trajectory}

In this section we derive the relativistic version of Newton's law for the
trajectory $\mathbf{\hat{r}}\left( t\right) $. We provide principal steps,
complete mathematical details presented in a similar situation can be found
in \cite{BF9}, \cite{BF10}. The equations are determined uniquely by the
assumptions made in the previous section.

We start with assigning the position to the distributed charge by using its
energy density $u=\mathcal{T}^{00}$. We define the \emph{energy center } $%
\mathbf{\mathbf{r}}\left( t\right) $ by the formula%
\begin{equation}
\mathbf{\mathbf{r}}\left( t\right) =\mathcal{\bar{E}}\left( t\right)
^{-1}\tint_{\Omega }\mathbf{\mathbf{x}}u\left( t,\mathbf{x}\right) \,\mathrm{%
d}\mathbf{x}  \label{enin}
\end{equation}%
where the concentrated energy $\mathcal{\bar{E}}\left( t\right) $ is defined
by (\ref{concen}). One can easily prove using (\ref{T00lim}) that 
\begin{equation}
\mathbf{\mathbf{r}}\left( t\right) \rightarrow \mathbf{\hat{r}}\left(
t\right) .  \label{rrlim}
\end{equation}

Using the conservation laws for the energy-momentum and the continuity
equation, we derive in Subsection \ref{Sconv} the following differential
equations for the trajectory and the energy $\mathcal{\bar{E}}_{\infty }$
concentrated at it: 
\begin{equation}
\partial _{t}\mathcal{\bar{E}}_{\infty }\left( t\right) =\hat{v}%
^{j}f_{\infty }^{j}+W_{\mathrm{rad}},  \label{enlim}
\end{equation}%
\begin{equation}
\partial _{t}\left( \mathcal{\bar{E}}_{\infty }\left( t\right) \hat{v}%
^{j}\right) =f_{\infty }^{j}+f_{\mathrm{rad}}^{j}.  \label{eqrad}
\end{equation}%
Here $\hat{v}^{j}=\partial _{t}\mathbf{\hat{r}}^{j}\left( t\right) $, the
Lorentz force $f_{\infty }^{j}$ is generated by the external field according
to formula (\ref{flin}), and $f_{\mathrm{rad}}^{j}$ is the radiation force
defined by (\ref{bound6r}). Obviously, (\ref{eqrad}) has the form of
Newton's law with the mass $M=\mathcal{\bar{E}}_{\infty }\left( t\right) $,
and this fact implies Einstein's formula on the equivalence of mass and
energy. Note that we derive this formula in a \ regime with acceleration
where Newtonian definition of mass is applicable. A similar derivation of
Einstein's formula in the regime without radiation is given in \cite{BF8}, 
\cite{BF9}. Examples of accelerating localized solutions of KG\ equations
are also given there. \ We will derive the LAD equation from equations (\ref%
{enlim}) and (\ref{eqrad}) in subsection \ref{SALD}.

\subsection{Rest mass of a charge and mass renormalization\label{Srest}}

In this section we define the rest mass as an integral of motion of the
system (\ref{enlim}), (\ref{eqrad}). \ We also demonstrate the mass
renormalization as the cancellation of two energies: the energy of the EM
field and the energy of the adjacent field.

We multiply equations (\ref{eqrad}) and (\ref{enlim}) \ by $\mathcal{\bar{E}}%
_{\infty }\hat{v}^{j}$\ and $\mathcal{\bar{E}}_{\infty }$\ respectively and
after subtraction obtain: 
\begin{equation}
\partial _{t}\left( \mathcal{\bar{E}}_{\infty }\left( t\right) \hat{v}%
^{j}\right) \mathcal{\bar{E}}_{\infty }\hat{v}^{j}-\mathcal{\bar{E}}_{\infty
}\partial _{t}\mathcal{\bar{E}}_{\infty }\left( t\right) =\mathcal{\bar{E}}%
_{\infty }\left( \hat{v}^{j}f_{\mathrm{rad}}^{j}-W_{\mathrm{rad}}\right) .
\label{dif1}
\end{equation}
Under the assumptions (\ref{gretc}), \ we obtain in Section \ref{sevint}
formula (\ref{rmc0}) which implies that the following radiative balance
condition is fulfilled:%
\begin{equation}
\hat{v}^{j}f_{\mathrm{rad}}^{j}-W_{\mathrm{rad}}=0.  \label{rmc}
\end{equation}%
Formulas (\ref{dif1}) and (\ref{rmc}) imply the following rest mass equation:

\begin{equation}
\partial _{t}\left[ \mathcal{\bar{E}}_{\infty }^{2}\left( t\right) \left(
1-\left\vert \mathbf{\hat{v}}\right\vert ^{2}\right) \right] =0.
\label{restm}
\end{equation}%
The expression in brackets equals a constant which we denote $M_{0}^{2}$,
and the constant $M_{0}$ is interpreted as the rest mass. Therefore, we
arrive to a familiar explicit expression for the time dependence of
concentrated energy: 
\begin{equation}
\mathcal{\bar{E}}_{\infty }\left( t\right) =\left( 1-\left\vert \mathbf{\hat{%
v}}\right\vert ^{2}\right) ^{-1/2}M_{0}.  \label{einm}
\end{equation}%
Equation (\ref{eqrad}) \ takes the form of the relativistic version of
Newton's law 
\begin{equation}
\partial _{t}\left( M_{0}\left( 1-\left\vert \mathbf{\hat{v}}\right\vert
^{2}\right) ^{-1/2}\hat{v}^{j}\right) =f_{\infty }^{j}+f_{\mathrm{rad}}^{j},
\label{reln}
\end{equation}%
where $M_{0}$ is the rest mass of the charge. \ \emph{The rest mass }$M_{0}$%
\emph{\ is the observable mass of the charge, and the existence of this
constant of motion signifies the mass renormalization. }Obviously, the rest
mass is not prescribed but emerges as an integral of motion of a system (\ref%
{enlim}), (\ref{eqrad}).

According to (\ref{uden}), the concentrated energy is given by the formula%
\begin{equation}
\mathcal{\bar{E}}_{\infty }=\lim_{R\rightarrow 0,\frac{a}{R}\rightarrow
0}\int_{\Omega }\left[ T^{00}+\Theta ^{00}\left( F\right) -\Theta
^{00}\left( F^{\mathrm{ad}}\right) \right] \mathrm{d}\mathbf{x}.
\label{Einfttt}
\end{equation}%
The tensor $\tilde{\Theta}^{00}=\Theta ^{00}\left( F\right) -\Theta
^{00}\left( F^{\mathrm{ad}}\right) $ \ which enters the above formula
quadratically depends on retarded and advanced fields $F_{\mathrm{ret}},F_{%
\mathrm{adv}}$. Note that the fields $F_{\mathrm{ret}}^{\mu \nu }$ and $F_{%
\mathrm{adv}}^{\mu \nu }$ derived from Li\'{e}nard--Wiechert potentials by
formula (\ref{Fmunug}) have singularities of order $\left\vert \mathbf{x-%
\hat{r}}\right\vert ^{-2}$ at the center of $\Omega $. \ The representation (%
\ref{Th00}) shows that $\tilde{\Theta}^{00}$ involves the radiation field $%
F_{\mathrm{rad}}=F_{\mathrm{ret}}-F_{\mathrm{adv}}$ as a factor. The crucial
property of the radiation field $F_{\mathrm{rad}}$ proved by Dirac \cite%
{Dirac CE} \ is its regularity. \ Thanks to the regularity, the tensor $%
\tilde{\Theta}^{00}$ has a mild singularity at the trajectory which is not
stronger than $\left\vert \mathbf{x-\hat{r}}\right\vert ^{-2}$ at the center
of $\Omega .$ Therefore it does not contribute to the limit in (\ref{Einfttt}%
) and 
\begin{equation}
\mathcal{\bar{E}}_{\infty }=\lim_{R\rightarrow 0,\frac{a}{R}\rightarrow
0}\int_{\Omega }T^{00}\mathrm{d}\mathbf{x}.  \label{Einft}
\end{equation}%
At the same time, both the EM energy and the energy of the adjacent field
which enter (\ref{Einfttt}) turn into infinity in the localization limit: 
\begin{equation*}
\int_{\Omega }\Theta ^{00}\left( F\right) \mathrm{d}\mathbf{x\rightarrow
\infty ,\quad }\int_{\Omega }\Theta ^{00}\left( F^{\mathrm{ad}}\right) 
\mathrm{d}\mathbf{x\rightarrow \infty }
\end{equation*}%
as $R\rightarrow 0,\frac{a}{R}\rightarrow 0$. \emph{Therefore, formulas (\ref%
{Einfttt}) and (\ref{Einft}) provide an explicit cancellation of the
infinite energies which constitutes an important part of the mass
renormalization.} In a contrast to the customary treatment of the mass
renormalization, \cite{Dirac CE}, \cite{LandauLif F}, \cite{Panofsky
Phillips}, \cite{Rohrlich}, \cite{Spohn}, \cite{Thirring}, where such a
cancellation in dynamical regimes is postulated, we demonstrate the
cancellation explicitly not as a result of a convenient subtraction of
unbounded terms, but as a result of calculation based on the specific
composition of the energy density which is uniquely determined in terms of
the Lagrangian of the problem.

Now we find the value of the rest mass $M_{0}$ in a regime which originates
from the uniform motion. If the motion is uniform for $t<0\ $with the
initial velocity $\mathbf{\hat{v}}_{\mathrm{init}}$ and the motion is
accelerated only for $t>0$, the value of the rest mass can be determined
based on the analysis of the uniform motion of a charge which is presented
in \cite{BF7}, \cite{BF8}, \cite{BF9}. \ Calculations made there show that 
\begin{equation}
\mathcal{\bar{E}}_{\infty }=M_{0}\left( 1-\left\vert \mathbf{\hat{v}}_{%
\mathrm{init}}\right\vert ^{2}\right) ^{-1/2},\quad M_{0}=m\mathrm{c}%
^{2}\left( 1+\Theta _{0}\frac{a_{\mathrm{C}}^{2}}{a^{2}}\right)
\label{Eptfree}
\end{equation}%
where $a_{\mathrm{C}}=\frac{\hbar }{m\mathrm{c}}=\frac{\hbar }{m}$ \ is the
reduced Compton wavelength \ and the factor $\Theta _{0}$ depends on the
nonlinearity. In particular, $\Theta _{0}=1/2$ \ for the logarithmic
nonlinearity (\ref{paf30}) corresponding to the Gaussian ground state (\ref%
{Gaus}). Note that if $\frac{a_{\mathrm{C}}}{a}$ converges to a non-zero
value, the nonlinearity provides a non-vanishing contribution to the rest
mass, and if $\frac{a_{\mathrm{C}}}{a}\rightarrow 0$ the rest mass $M_{0}=m%
\mathrm{\ \ }$coincides with the mass parameter in the KG\ equation. For a
discussion of the concept of mass see \cite{BF8}. The total charge $\bar{\rho%
}_{\infty }$ defined by (\ref{rhobar}), (\ref{paf6}) and (\ref{rhoinfg}) is
calculated in \cite{BF8}, \cite{BF9}, namely:

\begin{equation*}
\bar{\rho}_{\infty }=q
\end{equation*}%
where $q$ is the charge parameter in the KG equation defined by (\ref{KGone}%
), (\ref{comsemh}). For the electron we set $q=e$.

\subsection{Convergence in localization limit\label{Sconv}}

Now we proceed to the derivation of equations (\ref{enlim}) and (\ref{eqrad}%
).

\subsubsection{Charge convergence}

First we show that defined by (\ref{rhobar}) concentrated charge $\bar{\rho}%
_{n}$ \emph{converges to a constant}. \ We integrate the continuity
equation: 
\begin{equation}
\bar{\rho}_{n}\left( t\right) -\bar{\rho}_{n}\left( t_{0}\right)
=\int_{t_{0}}^{t}\int_{\partial \Omega }\left( \mathbf{\hat{v}}\cdot \mathbf{%
\bar{n}}\rho -\mathbf{\bar{n}}\cdot \mathbf{J}\,\right) \,\mathrm{d}\sigma
dt^{\prime }.  \label{intcon}
\end{equation}%
Since the integrals over the boundary converge to zero, we obtain that 
\begin{equation}
\bar{\rho}_{n}\left( t\right) \rightarrow \bar{\rho}_{\infty }\ \text{\ for}%
\ \ T_{-}<t<T_{+}  \label{rhoinfg}
\end{equation}%
where $\bar{\rho}_{\infty }$ does not depend on $t$. The above relation can
be written in the form 
\begin{equation}
\bar{J}^{0}=\bar{\rho}_{\infty }\hat{v}^{0}+o\left( 1\right)  \label{fijvh0}
\end{equation}%
where $\hat{v}^{0}=1$. \ 

\subsubsection{Lorentz force convergence}

Now we show that averaged Lorentz force density converges to the Lorentz
force acting on a point charge. \ \ Since $F^{\mu \nu }\left( t,\mathbf{x}%
\right) \,-F_{\infty }^{\mu \nu }\left( t\right) =o\left( 1\right) $ in $%
\Omega $, we conclude that 
\begin{gather}
\int_{\Omega }f^{\mu }\,\mathrm{d}\mathbf{x}=\int_{\Omega }F^{\mu \nu
}\,J_{\nu }\mathrm{d}\mathbf{x}  \label{fmu} \\
=\int_{\Omega }\,F_{\infty }^{\mu \nu }\left( t\right) J^{\nu }\mathrm{d}%
\mathbf{x+}\int_{\Omega }\left( F^{\mu \nu }\,-F_{\infty }^{\mu \nu }\right)
J_{\nu }\mathrm{d}\mathbf{x}=F_{\infty }^{\mu \nu }\left( t\right) \bar{J}%
_{\nu }+o\left( 1\right) .  \notag
\end{gather}%
Here 
\begin{equation}
\bar{J}^{\mu }=\int_{\Omega }J^{\mu }\mathrm{d}\mathbf{x;}  \label{Jbarmu}
\end{equation}%
for the component with $\mu =0$\ the above formula turns into (\ref{rhobar}%
). To find an expression for $\bar{J}^{i}$ \ we multiply the continuity
equation (\ref{paf6a})\ by the vector $\mathbf{x}-\mathbf{\hat{r}}$\ and
using the commutation relation $\partial _{j}\left( x^{i}\varphi \right)
-x^{i}\partial _{j}\varphi =\delta _{ij}\varphi $ we obtain the following
expression for $J^{i}$: 
\begin{equation}
\partial _{t}\left( \rho \left( x^{i}-\hat{r}^{i}\right) \right) +\partial
_{t}\hat{r}^{i}\rho +\partial _{j}\left( \left( x^{i}-\hat{r}^{i}\right)
J^{j}\right) =J^{i}\mathbf{,\quad }i=1,2,3.  \label{qPrg}
\end{equation}%
Integrating, we arrive to the equation 
\begin{equation}
\bar{J}^{i}=\hat{v}^{i}\bar{\rho}+\partial _{t}\int_{\Omega }\left( x^{i}-%
\hat{r}^{i}\right) \rho \mathrm{d}\mathbf{x}+\int_{\partial \Omega }\left(
x^{i}-\hat{r}^{i}\right) \left( \mathbf{\bar{n}}\cdot \mathbf{J}-\mathbf{%
\hat{v}\cdot \bar{n}}\rho \right) \,\mathrm{d}\sigma ,  \label{Javg}
\end{equation}%
where $\left( x^{i}-\hat{r}^{i}\right) =o\left( 1\right) $ in $\Omega $.
Together with (\ref{fijvh0}), the above equation implies that 
\begin{equation}
\bar{J}^{\nu }=\bar{\rho}_{\infty }\hat{v}^{\nu }+\partial _{t}o\left(
1\right) +o\left( 1\right) .  \label{Javga}
\end{equation}%
Therefore, (\ref{fmu}) implies an expression for the concentrated Lorentz
force density: 
\begin{equation}
\int_{\Omega }f^{\mu }\,\mathrm{d}\mathbf{x}=\bar{\rho}_{\infty }F_{\infty
}^{\mu \nu }\hat{v}_{\nu }+\partial _{t}o\left( 1\right) +o\left( 1\right) .
\label{fijvh1}
\end{equation}

We can write it in the form 
\begin{equation}
\int_{\Omega }f^{\mu }\,\mathrm{d}\mathbf{x}=f_{\infty }^{\mu }+o\left(
1\right) +\partial _{t}o\left( 1\right)  \label{lforl}
\end{equation}%
where the components of the Lorentz force $f_{\infty }^{j}$ are given by%
\begin{equation}
f_{\infty }^{j}=\bar{\rho}_{\infty }F_{\infty }^{j0}\left( t\right) +\bar{%
\rho}_{\infty }F_{\infty }^{ji}\left( t\right) \hat{v}_{i}  \label{flin}
\end{equation}%
and 
\begin{equation}
f_{\infty }^{0}=\bar{\rho}_{\infty }F_{\infty }^{0\nu }\hat{v}_{\nu }.
\label{flin0}
\end{equation}%
Note that since $F_{\infty }^{ji}$ is skew-symmetric, 
\begin{equation*}
f_{\infty }^{j}\hat{v}^{j}=\bar{\rho}_{\infty }F_{\infty }^{j0}\left(
t\right) \hat{v}^{j}+\bar{\rho}_{\infty }F_{\infty }^{ji}\left( t\right) 
\hat{v}_{i}\hat{v}^{j}=\bar{\rho}_{\infty }F_{\infty }^{j0}\left( t\right) 
\hat{v}^{j}=-\bar{\rho}_{\infty }F_{\infty }^{0\nu }\hat{v}^{\nu }
\end{equation*}%
and we can rewrite (\ref{flin0}) in the form 
\begin{equation}
f_{\infty }^{0}=\bar{\rho}_{\infty }F_{\infty }^{0\nu }\hat{v}_{\nu }=-\bar{%
\rho}_{\infty }F_{\infty }^{0\nu }\hat{v}^{\nu }=f_{\infty }^{j}\hat{v}^{j}.
\label{flin2}
\end{equation}

\subsubsection{Momentum convergence}

Multiplying the energy equation \ (\ref{equen2g}) by $\left( \mathbf{x}-%
\mathbf{r}\right) $, we obtain%
\begin{equation}
\partial _{t}\left( \left( \mathbf{x}-\mathbf{r}\right) u\right) +u\partial
_{t}\mathbf{r}=-\left( \mathbf{x}-\mathbf{r}\right) \partial
_{j}p^{j}+\left( \mathbf{x}-\mathbf{r}\right) f^{0}  \label{dtxmre0g}
\end{equation}%
where $\left( x^{i}-r^{i}\right) \partial _{j}p^{j}=\partial _{j}\left(
\left( x^{i}-r^{i}\right) p^{j}\right) -p^{i}$. We obtain from the relation (%
\ref{dtxmre0g}) an expression for $\mathcal{P}=p^{i}$ and its integral: 
\begin{gather}
\int_{\Omega }\mathcal{P}\mathrm{d}\mathbf{x}=\int_{\Omega }\partial
_{t}\left( \left( \mathbf{x}-\mathbf{r}\right) u\right) \mathrm{d}\mathbf{x}%
+\partial _{t}\mathbf{r}\int_{\Omega }u\mathrm{d}\mathbf{x}  \label{dtgrexh}
\\
+\int_{\partial \Omega }\left( \mathbf{x}-\mathbf{r}\right) n^{i}p^{i}%
\mathrm{d}\sigma -\int_{\Omega }\left( \mathbf{x}-\mathbf{r}\right) f^{0}%
\mathrm{d}\mathbf{x}.  \notag
\end{gather}%
We infer\ from the definition of the energy center $\mathbf{r}$ that%
\begin{equation}
\int_{\Omega }\partial _{t}\left( \left( \mathbf{x}-\mathbf{r}\right)
u\right) \,\mathrm{d}\mathbf{x}+\int_{\partial \Omega }\left( \mathbf{x}-%
\mathbf{r}\right) \mathbf{\hat{v}}\cdot \mathbf{\bar{n}}u\,\mathrm{d}\sigma
=0,  \label{dtxmreg}
\end{equation}%
therefore%
\begin{equation*}
\int_{\Omega }\partial _{t}\left( \left( \mathbf{x}-\mathbf{r}\right)
u\right) \,\mathrm{d}\mathbf{x}=o\left( 1\right) .
\end{equation*}%
Hence, taking into account that $\left( \mathbf{x}-\mathbf{r}\right)
=O\left( R\right) $ in $\Omega $, we conclude that 
\begin{equation}
\int_{\Omega _{n}}\mathcal{P}\mathrm{d}\mathbf{x}=\partial _{t}\mathbf{r}%
\mathcal{\bar{E}}+o\left( 1\right) .  \label{PQ02g}
\end{equation}

\subsubsection{Energy balance}

Integrating the energy equation (\ref{equen2g}) with respect to $x$ and\ $t$%
, we obtain the following equation:%
\begin{equation}
\mathcal{\bar{E}}\left( t\right) -\mathcal{\bar{E}}\left( t_{0}\right)
=\int_{t_{0}}^{t}\int_{\partial \Omega }\left( \mathbf{\hat{v}\cdot \bar{n}}%
u-\mathbf{\bar{n}}\cdot \mathcal{P}\right) \mathrm{d}\sigma \mathrm{d}%
t^{\prime }+\int_{t_{0}}^{t}\int_{\Omega }f^{0}\mathrm{d}\mathbf{x}\,\mathrm{%
d}t^{\prime }  \label{eneng}
\end{equation}%
where according to (\ref{lforl}) with $\mu =0$%
\begin{equation*}
\int_{t_{0}}^{t}\int_{\Omega }f^{0}\mathrm{d}\mathbf{x}\,\mathrm{d}t^{\prime
}=\int_{t_{0}}^{t}f_{\infty }^{0}\mathrm{d}t^{\prime }+o\left( 1\right) .
\end{equation*}%
Applying (\ref{bound3r}), we obtain an expression for the local energy
balance: 
\begin{equation}
\mathcal{\bar{E}}\left( t\right) -\mathcal{\bar{E}}\left( t_{0}\right)
=\int_{t_{0}}^{t}\,f_{\infty }^{0}\mathrm{d}t^{\prime }+\int_{t_{0}}^{t}W_{%
\mathrm{rad}}\mathrm{d}t^{\prime }+o\left( 1\right) .  \label{Einfr}
\end{equation}

\subsubsection{Momentum balance}

Integrating the momentum conservation equation (\ref{mom1g}) over $\Omega
\left( \mathbf{\hat{r}}(t),R\right) =\Omega $ and with respect to time, we
obtain%
\begin{gather}
\int_{\Omega }p^{j}\,\left( t,\mathbf{x}\right) \mathrm{d}\mathbf{x}%
-\int_{\Omega }p^{j}\left( t_{0},\mathbf{x}\right) \,\mathrm{d}\mathbf{x}%
-\int_{t_{0}}^{t}\int_{\Omega }f^{j}\mathrm{d}\mathbf{x}dt^{\prime }\,
\label{pint} \\
+\int_{t_{0}}^{t}\int_{\partial \Omega _{n}}\left( \bar{n}^{i}T^{ij}\,-\hat{v%
}^{i}\bar{n}^{i}p^{j}\right) \mathrm{d}\sigma \mathrm{d}t^{\prime }=0  \notag
\end{gather}%
where according to (\ref{lforl}) 
\begin{equation*}
\int_{t_{0}}^{t}\int_{\Omega }f^{j}\mathrm{d}\mathbf{x}dt^{\prime
}=\int_{t_{0}}^{t}f_{\infty }^{j}dt^{\prime }+o\left( 1\right) .
\end{equation*}%
\ Applying (\ref{PQ02g}) \ and (\ref{bound6r})\ we obtain 
\begin{equation}
\mathcal{\bar{E}}\left( t\right) \partial _{t}r^{j}\left( t\right) -\mathcal{%
\bar{E}}_{n}\left( t_{0}\right) \partial _{t}r^{j}\left( t_{0}\right)
=\int_{t_{0}}^{t}f_{\infty }^{j}dt^{\prime }+f_{\mathrm{rad}}^{j}+o\left(
1\right)  \label{dtern1ag}
\end{equation}%
with the Lorentz force given by (\ref{flin}).

\subsubsection{The necessary conditions on energy concentration}

Passing to the localization limit in \ (\ref{Einfr}), we obtain 
\begin{equation}
\mathcal{\bar{E}}_{\infty }\left( t\right) -\mathcal{\bar{E}}_{\infty
}\left( t_{0}\right) =\int_{t_{0}}^{t}f_{\infty }^{0}\,\mathrm{d}t^{\prime
}+\int_{t_{0}}^{t}W_{\mathrm{rad}}\mathrm{d}t^{\prime }.  \label{enintw}
\end{equation}%
Using (\ref{flin2}) we obtain equation (\ref{enlim}) for the limit energy.
Now we derive Newton's law (\ref{eqrad}). From (\ref{dtern1ag}), after
obvious manipulations, we obtain that 
\begin{equation*}
r^{j}\left( t\right) -r^{j}\left( t_{0}\right) -\int_{t_{0}}^{t}\frac{%
\mathcal{\bar{E}}\left( t_{0}\right) }{\mathcal{\bar{E}}\left( t\right) }%
\partial _{t}r^{j}\left( t_{0}\right) dt^{\prime \prime }=\int_{t_{0}}^{t}%
\frac{1}{\mathcal{\bar{E}}\left( t^{\prime \prime }\right) }%
\int_{t_{0}}^{t^{\prime \prime }}f_{\infty }^{j}dt^{\prime }dt^{\prime
\prime }+o\left( 1\right) .
\end{equation*}%
Using (\ref{rrlim}) and (\ref{T00lim}), we pass to the localization limit:%
\begin{equation*}
\hat{r}^{j}\left( t\right) -\hat{r}^{j}\left( t_{0}\right) -\partial _{t}%
\hat{r}^{j}\left( t_{0}\right) \int_{t_{0}}^{t}\frac{\mathcal{\bar{E}}%
_{\infty }\left( t_{0}\right) }{\mathcal{\bar{E}}_{\infty }\left( t\right) }%
dt^{\prime \prime }=\int_{t_{0}}^{t}\frac{1}{\mathcal{\bar{E}}_{\infty
}\left( t^{\prime \prime }\right) }\int_{t_{0}}^{t^{\prime \prime
}}f_{\infty }^{j}dt^{\prime }dt^{\prime \prime }.
\end{equation*}

From this integral equation we obtain differential equation (\ref{eqrad}).

\subsection{Radiation force and power for a localized charge}

To derive the LAD equation, we have to calculate the radiation power $W_{%
\mathrm{rad}}$ and radiation force $f_{\mathrm{rad}}^{j}$ in equations (\ref%
{rmc}) and (\ref{reln}). Note that since $\psi $ is localized, the KG\
tensor $T^{\mu \nu }$given by (\ref{TKG}) asymptotically vanish on $\partial
\Omega $ as $a/R\rightarrow 0$ and 
\begin{equation*}
\int_{\partial \Omega }\left( \hat{v}^{i}\bar{n}^{i}T^{00}-\bar{n}%
^{i}T^{0i}\right) \mathrm{d}\sigma =o\left( 1\right) .
\end{equation*}

Therefore definition (\ref{bound3r}), (\ref{bound6r}) can be rewritten in
the form 
\begin{equation}
W_{\mathrm{rad}}=\lim_{R\rightarrow 0}\int_{\partial \Omega }\left( \hat{v}%
^{i}\bar{n}^{i}\tilde{\Theta}^{00}-\bar{n}^{i}\tilde{\Theta}^{0i}\right) \,%
\mathrm{d}\sigma ,  \label{erad}
\end{equation}%
\begin{equation}
f_{\mathrm{rad}}^{j}=-\lim_{R\rightarrow 0}\int_{\partial \Omega }\left( 
\bar{n}^{i}\tilde{\Theta}^{ij}-\hat{v}^{i}\bar{n}^{i}\tilde{\Theta}%
^{0j}\right) \,\mathrm{d}\sigma .  \label{frad}
\end{equation}

The stress tensor $\tilde{\Theta}^{\mu \upsilon }$ is defined by formula (%
\ref{Tone}) with the tensor $\Theta ^{\mu \nu }\left( F\right) $ defined by (%
\ref{maxw11}) in terms of solutions of Maxwell equations (\ref{fmunu}), (\ref%
{fadmunu}). We assume that the solutions of the Maxwell equations are given
by the formulas (\ref{aj}), (\ref{aadj}) where $\mathcal{G}_{\mathrm{ret}}$
\ and $\mathcal{G}_{\mathrm{adv}}$ \ are the solution operators defined as
integral operators with retarded and advanced Green functions respectively,
\ in particular \ 
\begin{equation}
\mathcal{G}_{\mathrm{ret}}J^{\nu }\left( t,\mathbf{x}\right) =\dint \frac{%
\left[ J^{\nu }\left( t^{\prime },\mathbf{x}^{\prime }\right) \right] _{%
\mathrm{ret}}}{\left\vert \mathbf{x}-\mathbf{x}^{\prime }\right\vert }\,%
\mathrm{d}\mathbf{x}^{\prime },  \label{grmax12a}
\end{equation}%
where the symbol $\left[ \cdot \right] _{\mathrm{ret}}$ means that the
quantity in the square brackets is to be evaluated at the retarded time%
\begin{equation}
t^{\prime }=t_{\mathrm{ret}}=t-\left\vert \mathbf{x}-\mathbf{x}^{\prime
}\right\vert .  \label{grmax14}
\end{equation}

It is important to notice that the sources in equations (\ref{fmunu}) and (%
\ref{fadmunu}) are the same, and only the coefficients $\vartheta
_{0},\vartheta _{1}$ which determine the composition of Green functions are
different. Now we find the integrals (\ref{erad}), (\ref{frad}) under the
assumption (\ref{gretc}). \ Notice that \ the difference $F_{\mathrm{ret}%
}-F_{\mathrm{adv}}$ for the point sources does not have a singularity at $%
\mathbf{\hat{r}}\left( t\right) ,$\cite{Dirac CE}, and is a continuous
function.

The tensor $\Theta ^{\mu \nu }\left( F\right) $ defined by (\ref{maxw11}) is
quadratic with respect to field components $F^{\alpha \beta }$: $\Theta
^{\mu \nu }\left( F\right) =\Theta ^{\mu \nu }\left( FF\right) $. Taking
into account (\ref{aj}), (\ref{aadj}) we see that

\begin{equation}
\tilde{\Theta}^{\mu \nu }=\Theta ^{\mu \nu }\left( \left( F_{\mathrm{adv}%
}+\vartheta _{0}F_{\mathrm{rad}}\right) ^{2}\right) -\Theta ^{\mu \nu
}\left( \left( F_{\mathrm{adv}}+\vartheta _{1}F_{\mathrm{rad}}\right)
^{2}\right)  \label{ttild}
\end{equation}%
where%
\begin{equation}
F_{\mathrm{rad}}=F_{\mathrm{ret}}-F_{\mathrm{adv}}.  \label{Frad}
\end{equation}%
Expanding the quadratic tensors $\Theta ^{\mu \nu }$ we obtain 
\begin{equation}
\tilde{\Theta}^{\mu \nu }=\left( \vartheta _{0}-\vartheta _{1}\right) \tilde{%
\Theta}_{\mathrm{l}}^{\mu \nu }+\left( \vartheta _{0}^{2}-\vartheta
_{1}^{2}\right) \Theta ^{\mu \nu }\left( \left( F_{\mathrm{rad}}\right)
^{2}\right)  \label{Th00}
\end{equation}%
where%
\begin{equation}
\tilde{\Theta}_{\mathrm{l}}^{\mu \nu }=\Theta ^{\mu \nu }\left( F_{\mathrm{%
adv}}F_{\mathrm{rad}}\right) +\Theta ^{\mu \nu }\left( F_{\mathrm{rad}}F_{%
\mathrm{adv}}\right) .  \label{Th0}
\end{equation}

Since $F_{\mathrm{rad}}\ $ has no singularity at $\mathbf{\hat{r}}\left(
t\right) $, $\ $the integral of $\Theta ^{\mu \nu }\left( \left( F_{\mathrm{%
rad}}\right) ^{2}\right) $ \ gives a vanishing contribution \ as $%
R\rightarrow 0$ and we can replace $\tilde{\Theta}^{\mu \nu }$ by $\tilde{%
\Theta}_{\mathrm{l}}^{\mu \nu }$ in (\ref{erad}), (\ref{frad}). \ Note that
formula (\ref{Th0}) \ \ already indicates that limits (\ref{erad}), (\ref%
{frad}) are well-defined since the only singularity comes from $F_{\mathrm{%
adv}}$ and it has the leading order $R^{-2}$, whereas the sphere area is $%
4\pi R^{2}$. \ But still we need to evaluate the integrals involving $\tilde{%
\Theta}_{\mathrm{l}}^{\mu \nu }$.

To evaluate $\tilde{\Theta}_{\mathrm{l}}^{\mu \nu },$ we use expressions
obtained by Dirac \cite{Dirac CE} for the fields $F_{\mathrm{rad}}=F_{%
\mathrm{ret}}-F_{\mathrm{adv}}$, $F_{\mathrm{ret}}$ and $F_{\mathrm{adv}}$.
The expressions are given in the following section.

\subsection{Energy momentum tensors of EM field near trajectory}

Dirac \cite{Dirac CE} obtained expressions for the EnMT of the field
generated by a point charge. We have to note that Dirac used notation $%
A_{\nu }$ for the EM potential whereas we use notation $A^{\nu }$ for the
same potential and the tensor $F^{\mu \nu }$ in Dirac's notation equals $%
F_{\mu \nu }$ in our notation, hence we rewrite the formulas in our
notation. We also denote by $\left( vw\right) $ the scalar product of
4-vectors given by (\ref{vw}), $\left( vw\right) =v_{0}w_{0}-\mathbf{v\cdot w%
}$ where $\mathbf{v\cdot w}$ is the usual dot product in 3D space. We assume
a parametrization of the trajectory 
\begin{equation*}
\left( t,\mathbf{\hat{r}}(t)\right) =\left( r^{0},\mathbf{\hat{r}}(t)\right)
=z\left( s\right) =\left( z_{0}\left( s\right) ,\mathbf{z}\left( s\right)
\right)
\end{equation*}%
by the proper time $s$ so that 
\begin{equation*}
v^{\alpha }=\partial _{s}r^{\alpha }=\dot{z}^{\alpha }\left( s\right)
=\left( v^{0},\mathbf{v}\right)
\end{equation*}%
satisfies the normalization 
\begin{equation*}
\left( vv\right) =v_{0}^{2}-\left\vert \mathbf{v}^{2}\right\vert =1.
\end{equation*}%
The normalization implies identities 
\begin{equation*}
\left( v\dot{v}\right) =0,\quad \left( v\ddot{v}\right) +\left( \dot{v}\dot{v%
}\right) =0
\end{equation*}%
where $\dot{v}=\partial _{s}v$. Note that 
\begin{equation*}
\hat{v}_{i}=v_{i}/v_{0}.
\end{equation*}

We fix a point of a trajectory and a sphere $\partial \Omega $ of radius $R$
centered at it. We need to find the values of the tensors $\tilde{\Theta}_{%
\mathrm{l}}^{\mu \nu }$ on the sphere $\partial \Omega $. We choose the
coordinates so that the center of the sphere is at the origin. For the
calculation, we also assume that $s=0$ corresponds to the center of the
sphere. \ Therefore a point $x=\left( t,\mathbf{x}\right) $ on the sphere $%
\partial \Omega $ satisfies the equation $t=0,\left\vert \mathbf{x}%
\right\vert =R$ and the external normal to the sphere can be written as $%
\mathbf{\bar{n}}=\frac{\mathbf{x}}{R}$. Taking into account (\ref{Th00}), we
can write (\ref{erad}), (\ref{frad}) in the form \ 
\begin{equation}
W_{\mathrm{rad}}=\vartheta \lim_{R\rightarrow 0}\frac{1}{R}\int_{\partial
\Omega }\left( \frac{v^{i}}{v_{0}}x^{i}\tilde{\Theta}_{\mathrm{ll}%
}^{00}-x^{i}\tilde{\Theta}_{\mathrm{ll}}^{0i}\right) \,\mathrm{d}\sigma ,
\label{erad0}
\end{equation}%
\begin{equation}
f_{\mathrm{rad}}^{j}=-\vartheta \lim_{R\rightarrow 0}\frac{1}{R}%
\int_{\partial \Omega }\left( x^{i}\tilde{\Theta}_{\mathrm{ll}}^{ij}-\frac{%
v^{i}}{v_{0}}x^{i}\tilde{\Theta}_{\mathrm{ll}}^{0j}\right) \,\mathrm{d}%
\sigma ,  \label{frad0}
\end{equation}%
where $\vartheta =\vartheta _{0}-\vartheta _{1},$and $\tilde{\Theta}_{%
\mathrm{ll}}^{\mu \nu }$ is the leading part of the tensor (\ref{Th0}):%
\begin{equation*}
\tilde{\Theta}_{\mathrm{l}}^{\mu \nu }=\tilde{\Theta}_{\mathrm{ll}}^{\mu \nu
}+o\left( R^{-2}\right) .
\end{equation*}

The tensor $\tilde{\Theta}_{\mathrm{ll}}^{\mu \nu }$ is explicitly written
below in (\ref{Thlead}), and now we describe its derivation. According to (%
\ref{Th0}), $\tilde{\Theta}_{\mathrm{l}}^{\mu \nu }$ involves products of
components of $F_{\mathrm{rad}}$ and $F_{\mathrm{adv}}$. \ Now we write the
leading terms of the expressions for $F_{\mathrm{rad}}$ and $F_{\mathrm{adv}%
} $ from \cite{Dirac CE} which contribute to the limits (\ref{erad}), (\ref%
{frad}). The $\ $field $F_{\mathrm{rad}\mu \upsilon }$ has no singularity,
hence the leading terms for $F_{\mathrm{rad}\mu \upsilon }$ are of zero
order in $R$, and the leading terms for $F_{\mathrm{adv}\mu \upsilon }$ are
of order $R^{-2}$. The leading term of formula (12) in \cite{Dirac CE} takes
the form 
\begin{equation}
F_{\mathrm{rad}\mu \upsilon }=\frac{4e}{3}\left( \ddot{v}^{\mu }v^{\upsilon
}-\ddot{v}^{\upsilon }v^{\mu }\right)  \label{radf1}
\end{equation}%
where $e$ is the value of the point charge. The leading part $F_{\mathrm{0}%
\mu \upsilon }$ of the tensor $F_{\mathrm{adv}\mu \upsilon }$ (which
coincides with the leading part of $F_{\mathrm{ret}\mu \upsilon }$) is given
by the expression 
\begin{equation}
F_{\mathrm{0}\mu \upsilon }=-e\epsilon ^{-3}\left( v^{\mu }\gamma ^{\upsilon
}-v^{\upsilon }\gamma ^{\mu }\right)  \label{f0}
\end{equation}%
obtained from formula (60) in \cite{Dirac CE}. Now we explain the notation
in this formula. The 4-vector $\gamma =\gamma ^{\upsilon }$ is defined by
the relation 
\begin{equation*}
x^{\mu }=z^{\mu }\left( s_{0}\right) +\gamma ^{\mu }
\end{equation*}%
where $s_{0}$ \ is such that 
\begin{equation*}
\left( \gamma \dot{z}\left( s_{0}\right) \right) =0.
\end{equation*}%
The 4-vector $\gamma $ is space-like: 
\begin{equation*}
-\epsilon ^{2}=\left( \gamma \gamma \right) =\gamma _{0}^{2}-\mathbf{\gamma }%
^{2}
\end{equation*}%
where $\epsilon >0$ is the same as in (\ref{f0}).

Obviously $\gamma =O\left( R\right) $ on $\partial \Omega $, and, to find
the leading term, it is sufficient to find $s_{0}$ and $\gamma $ with the
accuracy $o\left( R\right) $. The trajectory can be approximated by the
tangent straight line with the accuracy $O\left( R^{2}\right) $, and we can
determine $s_{0}$ using this approximation. An elementary calculation
produces 
\begin{equation}
s_{0}=x_{0}\dot{z}_{0}-\mathbf{x\cdot v}  \label{s0x}
\end{equation}%
where $\mathbf{v=\dot{z}}$ \ and 
\begin{equation}
\gamma =x-\dot{z}\left( x_{0}\dot{z}_{0}-\mathbf{x\cdot v}\right) =x-\left( x%
\dot{z}\right) \dot{z},  \label{gams}
\end{equation}%
$x_{0}=0.$ Calculating $\left( \gamma \gamma \right) $ we obtain 
\begin{equation}
\epsilon =\left( R^{2}+\left( \mathbf{x\cdot v}\right) ^{2}\right) ^{1/2}.
\label{epsxz}
\end{equation}%
Therefore all the terms in (\ref{f0}) are explicitly expressed in terms of $%
\mathbf{x}$ and the derivatives of trajectory coordinates. The derivatives
have to be calculated at $s=s_{0}$, but for the leading term we can take
their values at $s=0$ that is at the center of the sphere $\partial \Omega $.

We have to evaluate $\tilde{\Theta}_{\mathrm{l}}^{\mu \nu }$ defined by (\ref%
{Th0}) where $\Theta ^{\mu \nu }\left( F\right) $ is defined by (\ref{maxw11}%
). A straightforward computation \ produces 
\begin{equation}
\tilde{\Theta}_{\mathrm{l}}^{\mu \nu }=\frac{1}{4\pi }g^{\mu \alpha }F_{%
\mathrm{0}\alpha \beta }F_{\mathrm{rad}}^{\beta \nu }+\frac{1}{4\pi }g^{\mu
\alpha }F_{\mathrm{rad}\alpha \beta }F_{\mathrm{0}}^{\beta \nu }+\frac{1}{%
8\pi }g^{\mu \nu }F_{\mathrm{0}\alpha \beta }F_{\mathrm{rad}}^{\alpha \beta
}.  \label{Thlead}
\end{equation}

Now we substitute expressions (\ref{radf1}) and (\ref{f0}) into the above
formula. After straightforward calculations we obtain the following
expression for the leading term of $\tilde{\Theta}_{\mathrm{l}}^{\mu \nu }$
on $\partial \Omega $: 
\begin{eqnarray}
-\frac{3\pi }{e^{2}\epsilon ^{-3}}\tilde{\Theta}_{\mathrm{ll}}^{\mu \nu } &=&%
\left[ 2v_{\mu }v_{\upsilon }\ddot{v}_{\beta }-v_{\beta }\left( v_{\upsilon }%
\ddot{v}_{\mu }+v_{\mu }\ddot{v}_{\upsilon }\right) +g^{\mu \nu }\left(
\left( v\ddot{v}\right) v_{\beta }-\ddot{v}_{\beta }\right) \right] x^{\beta
}  \notag \\
&&+x_{\mu }\left( \ddot{v}_{\upsilon }-\left( v\ddot{v}\right) v_{\upsilon
}\right) +x_{\upsilon }\left( \ddot{v}_{\mu }-\left( v\ddot{v}\right) v_{\mu
}\right)  \label{Thmunuc}
\end{eqnarray}%
with $x_{0}=0$ on $\partial \Omega $ .

\subsubsection{Evaluation of the integrals\label{sevint}}

According to (\ref{Thmunuc}), the integrals (\ref{erad0}), (\ref{frad0})
over the sphere $\partial \Omega $ involve integrals of the form 
\begin{gather}
\Xi \left( x^{i},x^{j}\right) =\Xi \left( x_{i},x_{j}\right) =\frac{1}{R}%
\int_{\partial \Omega }x_{i}x_{j}\epsilon ^{-3}\mathrm{d}\sigma
\label{Ksixx0} \\
=\frac{1}{R}\int_{\left\vert \mathbf{x}\right\vert =R}x_{i}x_{j}\left(
R^{2}+\left( \mathbf{x\cdot v}\right) ^{2}\right) ^{-3/2}\mathrm{d}\sigma . 
\notag
\end{gather}%
Evaluating the above integral (see Appendix), we obtain that 
\begin{equation}
\Xi \left( x_{i},x_{j}\right) =\frac{v_{i}v_{j}}{\mathbf{v}^{2}}\left[ \Xi
_{\Vert }-\Xi _{\bot }\right] +\delta _{ij}\Xi _{\bot }.  \label{ksid}
\end{equation}%
The \ coefficients $\Xi _{\Vert }$ and $\Xi _{\bot }$, which depend only on $%
\left\vert \mathbf{v}\right\vert $, are calculated in the Appendix, where
the following useful formula is obtained: 
\begin{equation}
\Xi _{\Vert }+2\Xi _{\bot }=\frac{4\pi }{\sqrt{\left\vert \mathbf{v}%
\right\vert ^{2}+1}}=\frac{4\pi }{v_{0}}.  \label{ksipl}
\end{equation}

When we use formula (\ref{ksid}), it is convenient to evaluate separately
terms at coefficients $\left[ \Xi _{\Vert }-\Xi _{\bot }\right] $ and $\Xi
_{\bot }$ . After elementary but tedious calculations, we obtain expressions
for the integrals in (\ref{erad0}), (\ref{frad0}): 
\begin{gather}
-\frac{3\pi }{e^{2}}\frac{1}{R}\int x^{i}\tilde{\Theta}_{\mathrm{ll}%
}^{ij}\epsilon ^{-3}\mathrm{d}\sigma =\left[ \Xi _{\Vert }-\Xi _{\bot }%
\right] \left[ v^{j}\mathbf{\ddot{v}\cdot v}-\mathbf{v}^{2}\ddot{v}%
^{j}-\left( \ddot{v}^{j}-v^{j}\left( v\ddot{v}\right) \right) \right]
\label{one} \\
+\Xi _{\bot }\left[ \mathbf{\ddot{v}\cdot v}v^{j}-\mathbf{v}^{2}\ddot{v}%
^{j}-3\ddot{v}^{j}+3v^{j}\left( v\ddot{v}\right) \right] ,  \notag
\end{gather}%
\begin{equation}
-\frac{3\pi }{e^{2}}\frac{1}{R}\int \hat{v}^{i}x^{i}\tilde{\Theta}_{\mathrm{%
ll}}^{0j}\epsilon ^{-3}\mathrm{d}\sigma =\Xi _{\Vert }\left[ -\mathbf{\ddot{v%
}\cdot v}v^{j}+\mathbf{v}^{2}\ddot{v}^{j}\right] ,  \label{one2}
\end{equation}%
\begin{equation}
-\frac{3\pi }{e^{2}}\frac{1}{R}\int \tilde{\Theta}_{\mathrm{ll}%
}^{0j}x^{j}\epsilon ^{-3}\mathrm{d}\sigma =-2\Xi _{\bot }\left( \mathbf{v}%
^{2}\ddot{v}_{0}-v_{0}\mathbf{\ddot{v}\cdot v}\right) ,  \label{one3}
\end{equation}%
\begin{equation}
-\frac{3\pi }{e^{2}}R^{-1}\frac{1}{v_{0}}\int \left( v^{i}x^{i}\Theta _{%
\mathrm{ll}}^{00}\right) \epsilon ^{-3}\mathrm{d}\sigma =\Xi _{\Vert }\left(
-v_{0}\mathbf{\ddot{v}\cdot v}+\mathbf{v}^{2}\ddot{v}_{0}\right) .
\label{one4}
\end{equation}

Substituting the above expressions, we obtain 
\begin{equation}
W_{\mathrm{rad}}=-\frac{e^{2}}{3\pi }\vartheta \left( \Xi _{\Vert }+2\Xi
_{\bot }\right) \left( -v_{0}\mathbf{\ddot{v}\cdot v}+\mathbf{v}^{2}\ddot{v}%
_{0}\right) ,  \label{Epr0}
\end{equation}%
\begin{equation}
f_{\mathrm{rad}}^{j}=\frac{e^{2}}{3\pi }\vartheta \left( \Xi _{\Vert }+2\Xi
_{\bot }\right) \left[ \ddot{v}^{j}-v^{j}\left( v\ddot{v}\right) \right] .
\label{fradj}
\end{equation}%
Now we are able to check the radiative balance condition (\ref{rmc}):%
\begin{gather}
v_{0}W_{\mathrm{rad}}-\mathbf{v\cdot f}_{\mathrm{rad}}  \label{rmc0} \\
=-\frac{e^{2}}{3\pi }\vartheta \left( \Xi _{\Vert }+2\Xi _{\bot }\right) %
\left[ -v_{0}^{2}\mathbf{\ddot{v}\cdot v}+\mathbf{\ddot{v}\cdot v}+\mathbf{v}%
^{2}\mathbf{\ddot{v}\cdot v}\right] =0.  \notag
\end{gather}%
Using (\ref{ksipl}), we rewrite the radiative force as follows: 
\begin{equation}
f_{\mathrm{rad}}^{j}=\frac{4e^{2}}{3v_{0}}\vartheta \left[ \ddot{v}%
^{j}+v^{j}\left( \dot{v}\dot{v}\right) \right] .  \label{fradj0}
\end{equation}

\subsubsection{Generalized LAD equation\label{SALD}}

Rewriting (\ref{reln}) in terms of the proper time $s,$ we obtain

\begin{equation}
\partial _{s}v^{j}=v_{0}f_{\infty }^{j}+\frac{4e^{2}}{3}\vartheta \left[ 
\ddot{v}^{j}+v^{j}\left( \dot{v}\dot{v}\right) \right] .  \label{dsv}
\end{equation}

Equation (\ref{enlim}), where according to (\ref{flin2}) $\hat{v}%
^{j}f_{\infty }^{j}=\bar{\rho}_{\infty }F_{\infty }^{0\nu }\hat{v}_{\nu
}=v_{\mu }\bar{\rho}_{\infty }F_{\infty }^{0\mu }/v^{0}$, \ \ takes the form 
\begin{gather}
\partial _{s}\left( v_{0}M_{0}\right) =v_{0}v^{j}f_{\infty }^{j}+v_{0}W_{%
\mathrm{rad}}=v_{\mu }\bar{\rho}_{\infty }F_{\infty }^{0\mu }  \label{dsv1}
\\
-\frac{4e^{2}\vartheta }{3v_{0}}\left( -v_{0}^{2}\mathbf{\ddot{v}\cdot v}+%
\mathbf{v}^{2}\ddot{v}_{0}v_{0}\right) =v_{\mu }\bar{\rho}_{\infty
}F_{\infty }^{0\mu }+\frac{4e^{2}}{3}\vartheta \left( \ddot{v}%
_{0}-v_{0}\left( \ddot{v}v\right) \right) .  \notag
\end{gather}%
Taking (\ref{dsv}) and (\ref{dsv1}) together, we obtain 
\begin{equation*}
\partial _{s}v^{\mu }=F_{\infty }^{\mu \nu }\left( z\left( s\right) \right)
v_{\nu }+\frac{4e^{2}}{3}\vartheta \left[ \ddot{v}^{\mu }+v^{\mu }\left( 
\dot{v}\dot{v}\right) \right]
\end{equation*}%
implying (\ref{ALDeqt}).

\section{Many interacting charges\label{Smany}}

When we consider a single charge, the EM field $A^{\mu }$ seems quite
similar to the adjacent field $A^{\mathrm{ad}\mu }.$ To see the difference,
we have to consider many interacting charges, \ and one can see from this
construction that the EM field potential $A^{\mu }$ is responsible for the
interaction between the charges and is an observable field, whereas $A^{%
\mathrm{ad}\mu }$ \ is solely responsible for an internal interaction of a
charge with itself.

We consider a closed system of $N$ charges with densities $\psi _{\ell },$ $%
\ell =1,...,N$; the charges interact through the electromagnetic field $%
A=A^{\mu }$ \ and every charge interacts also with its adjacent field $%
A_{\ell }^{\mathrm{ad}}$, and the external EM\ field is absent: $A_{\mathrm{%
ex}}^{\mu }=0.$

The Lagrangian for the system of interacting charges has the form 
\begin{equation}
\mathcal{L}=\dsum\nolimits_{\ell }L_{\mathrm{KG}}\left( \psi _{\ell },\psi
_{\ell ;\mu },\psi _{\ell }^{\ast },\psi _{\ell ;\mu }^{\ast }\right) -L_{%
\mathrm{em}}\left( A\right) +\dsum\nolimits_{\ell }L_{\mathrm{em}}\left(
A_{\ell }^{\mathrm{ad}}\right) ,  \label{flagr6a}
\end{equation}%
where the EM\ Lagrangian $L_{\mathrm{em}}$ is defined by (\ref{Lem}), and $%
L_{\mathrm{KG}}$ is the Lagrangian (\ref{paf1h}) of the nonlinear
Klein-Gordon (KG) equation. It involves covariant derivatives 
\begin{equation}
\psi _{\ell }^{;\mu }=\tilde{\partial}_{\ell }^{\mu }\psi _{\ell }=\partial
^{\mu }\psi _{\ell }+\frac{\mathrm{i}q_{\ell }}{\hbar }\tilde{A}_{\ell
}^{\mu }\psi _{\ell },  \label{comsem}
\end{equation}%
(we do not sum over repeated $\ell $)%
\begin{equation}
\tilde{\partial}_{\ell }^{\mu }=\partial ^{\mu }+\frac{\mathrm{i}q_{\ell }}{%
\hbar }\tilde{A}_{\ell }^{\mu },\ \ \partial ^{\mu }=\frac{\partial }{%
\partial x_{\mu }}=\left( \partial _{t},-\nabla \right) ,  \label{covder}
\end{equation}%
\begin{equation}
\tilde{A}_{\ell }^{\mu }=A^{\mu }-A_{\ell }^{\mathrm{ad}\mu }.  \label{Aell}
\end{equation}%
In the simplest case which we consider here, $\psi _{\ell }$ is a complex
scalar, $\psi _{\ell }^{\ast }$ is complex conjugate to $\psi _{\ell }%
\mathrm{.}$\ The field equations which describe dynamics of the system are
obtained as Euler-Lagrange equations for the Lagrangian $\mathcal{L}$ and
involve the nonlinear Klein-Gordon (KG) equations for the charge
distributions $\psi _{\ell }$ and Maxwell equations for the potentials $%
A,A_{\ell }^{\mathrm{ad}}$. The KG equations have the form 
\begin{equation}
\left[ \tilde{\partial}_{\ell }^{\mu }\tilde{\partial}_{\ell }^{\mu }+\kappa
_{0}^{2}+G^{\prime }\left( \psi _{\ell }^{\ast }\psi _{\ell }\right) \right]
\psi _{\ell }=0.  \label{KGL}
\end{equation}%
Equations for the fields $F^{\mu \nu },F_{\ell }^{\mathrm{ad}\mu \nu }$ take
the form \ of Maxwell equations: 
\begin{equation}
\partial _{\mu }F^{\mu \nu }=4\pi J^{\nu },\quad J^{\nu
}=\dsum\nolimits_{\ell }J_{\ell }^{\nu }  \label{flagr8d}
\end{equation}%
\begin{equation}
\partial _{\mu }F_{\ell }^{\mathrm{ad}\mu \nu }=4\pi J_{\ell }^{\nu },
\label{flagr8dd}
\end{equation}%
where the source currents are defined by%
\begin{equation}
J_{\ell }^{\nu }=-\frac{\partial L_{\ell }}{\partial A_{\nu }}=-\mathrm{i}%
\frac{q_{\ell }}{\hbar }\left( \frac{\partial L_{\ell }}{\partial \psi
_{\ell ;\nu }}\psi _{\ell }-\frac{\partial L_{\ell }}{\partial \psi _{\ell
;\nu }^{\ast }}\psi _{\ell }^{\ast }\right) .  \label{Jmax}
\end{equation}%
Every current $J_{\ell }^{\nu }$ satisfies the continuity equation,
therefore we impose Lorentz gauge on solutions of (\ref{flagr8d}), (\ref%
{flagr8dd}). We choose the solution operator so that a solution of (\ref%
{flagr8d}) is given by (\ref{aj}), and similarly to (\ref{aadj}) $A_{\ell }^{%
\mathrm{ad}\nu }=\vartheta _{\ell }\mathcal{G}_{\mathrm{ret}}J^{\nu }+\left(
1-\vartheta _{\ell }\right) \mathcal{G}_{\mathrm{adv}}J^{\nu }$. Since the
Maxwell equation is linear, the total EM field is a superposition of the
fields for a single charge:%
\begin{equation*}
A^{\mu }=\dsum\nolimits_{\ell }A_{\ell }^{\mu },\quad F^{\mu \nu
}=\dsum\nolimits_{\ell }F_{\ell }^{\mu \nu }
\end{equation*}%
with the following equation for every field: 
\begin{equation}
\partial _{\mu }F_{\ell }^{\mu \nu }=4\pi J_{\ell }^{\nu }.  \label{maxl}
\end{equation}

It is important to note that the KG equation for $\ \ell $-th charge
involves (through the covariant derivative) the electromagnetic potential 
\begin{equation}
\tilde{A}_{\ell }^{\mu }=A_{\neq \ell }^{\mu }+A_{\ell }^{\mu }-A_{\ell }^{%
\mathrm{ad}\mu }
\end{equation}%
where 
\begin{equation*}
A_{\neq \ell }^{\mu }=\dsum\nolimits_{\ell ^{\prime }\neq \ell }A_{\ell
^{\prime }}^{\mu }.
\end{equation*}%
Comparing with (\ref{aaex}), we see that the potential $A_{\neq \ell }^{\mu
} $ \ for $\ell $-th charge plays the role of an external field $A_{\mathrm{%
ex}}^{\mu }$ for this charge. If we add a test charge to the system, it
becomes $N+1$-st charge and the field potential $\ A_{\neq N+1}^{\mu }$
coincides with the total field $A^{\mu }$ determined from (\ref{flagr8d}).
Therefore the observable field is the field $F^{\mu \nu }$ with the
potential $A^{\mu } $. \ Note that none of the adjacent potentials $A_{\ell
}^{\mathrm{ad}\mu },$ $\ell =1,...,N,$ enters the equations which determine
the dynamics of the test charge. Therefore $A_{\ell }^{\mathrm{ad}\mu }$
describes solely an internal self-interaction of a charge and this property
completely differs the adjacent field from the EM\ field which acts on all
charges of the system.

\section{Appendix: elementary surface integrals}

The integral of the form (\ref{Ksixx0}) determines a bilinear form on linear
functions $\mathbf{f}$ on the 3D space: 
\begin{equation*}
\Xi \left( \mathbf{f}_{1},\mathbf{f}_{2}\right) =\int_{\left\vert \mathbf{x}%
\right\vert =1}\frac{\mathbf{f}_{1}\mathbf{f}_{2}}{\left( R^{2}+\left( 
\mathbf{xv}\right) ^{2}\right) ^{3/2}}\mathrm{d}\sigma .
\end{equation*}%
As a first step, we evaluate the form for special cases. If we take the axis 
$x_{3}$ axis along $\mathbf{v}$ we obtain $\mathbf{v}=\left(
0,0,v^{3}\right) ,\ $\ $\left\vert \mathbf{v}\right\vert =\left\vert
v_{3}\right\vert $ \ and we consider 
\begin{equation*}
\int x_{i}x_{j}\epsilon ^{-3}d\sigma =\int \frac{x_{i}x_{j}}{\left(
R^{2}+\left( x_{3}^{2}v_{3}^{2}\right) \right) ^{3/2}}d\sigma .
\end{equation*}%
Since $x_{i}x_{j}$ is odd with respect to $x_{i}$ for $i\neq j$, \ we obtain 
\begin{equation}
\frac{1}{R}\int x_{i}x_{j}\left( R^{2}+\left( x_{3}^{2}v_{3}^{2}\right)
\right) ^{-3/2}d\sigma =0\text{ \ for }i\neq j.  \label{xinej}
\end{equation}%
In spherical coordinates $\ x_{3}=r\cos \theta ,x_{1}=r\sin \theta \cos
\varphi ,x_{2}=r\sin \theta \sin \varphi $ and 
\begin{equation*}
\Xi \left( x_{i},x_{j}\right) =R^{-1}\int_{-\pi }^{\pi }\int_{0}^{\pi }\frac{%
x_{i}x_{j}}{\left( R^{2}+x_{3}^{2}v_{3}^{2}\right) ^{3/2}}r^{2}\sin \theta
d\theta d\varphi .
\end{equation*}%
When $\mathbf{f}$ is parallel to $\mathbf{v}$, we set $i=j=3$: 
\begin{gather*}
\frac{1}{R}\int_{-\pi }^{\pi }\int_{0}^{\pi }\frac{x_{3}^{2}}{\left(
R^{2}+x_{3}^{2}v_{3}^{2}\right) ^{3/2}}r^{2}\sin \theta d\theta d\varphi \\
=4\pi \int_{0}^{1}\frac{t^{2}\ }{\left( 1+t^{2}v_{3}^{2}\right) ^{3/2}}%
dt=\Xi _{\Vert }\left( \left\vert v_{3}\right\vert \right)
\end{gather*}%
where 
\begin{equation*}
\Xi _{\Vert }\left( b\right) =-\frac{4\pi }{b^{3}}\frac{b-\left( \ln \left(
b+\sqrt{b^{2}+1}\right) \right) \sqrt{b^{2}+1}}{\sqrt{b^{2}+1}}.
\end{equation*}%
Therefore 
\begin{equation*}
\Xi \left( \frac{\mathbf{v}}{\left\vert \mathbf{v}\right\vert },\frac{%
\mathbf{v}}{\left\vert \mathbf{v}\right\vert }\right) =\int_{\Omega
}x_{3}^{2}\left( R^{2}+\left( x_{3}^{2}v_{3}^{2}\right) \right)
^{-3/2}d\sigma =\Xi _{\Vert }\left( \left\vert \mathbf{v}\right\vert \right)
.
\end{equation*}%
For the orthogonal to $\mathbf{v}$ direction $\mathbf{f}$, we evaluate%
\begin{equation*}
\frac{1}{R}\int_{-\pi }^{\pi }\int_{0}^{\pi }\frac{x_{1}^{2}}{\left(
R^{2}+x_{3}^{2}v_{3}^{2}\right) ^{3/2}}r^{2}\sin \theta d\theta d\varphi
=2\pi \int_{0}^{1}\frac{1\ -t^{2}}{\left( 1+t^{2}v_{3}^{2}\right) ^{3/2}}%
dt=\Xi _{\bot }\left( \left\vert v_{3}\right\vert \right)
\end{equation*}%
where%
\begin{equation*}
\Xi _{\bot }\left( \left\vert \mathbf{v}\right\vert \right) =2\pi \frac{1}{%
\sqrt{\left\vert \mathbf{v}\right\vert ^{2}+1}}-\frac{1}{2}\Xi _{\Vert
}\left( \left\vert \mathbf{v}\right\vert \right) .
\end{equation*}%
Therefore, if \ $\mathbf{f}$ is orthogonal to $\mathbf{v}$, namely $\mathbf{%
f\cdot v}=\mathbf{0}$ and $\ \left\vert \mathbf{f}\right\vert =1,$ we have 
\begin{equation*}
\ \Xi \left( \mathbf{f},\mathbf{f}\right) =\Xi _{\bot }\left( \left\vert 
\mathbf{v}\right\vert \right) .
\end{equation*}

Now we consider a general functional $\mathbf{f}$\ \ \ on $\mathbb{R}^{3}$,
the functional can be written in the form $\mathbf{f}=f_{i}x_{i}$ and we
define its norm $\left\vert \mathbf{f}\right\vert =\left( f_{i}f_{i}\right)
^{1/2}$. The functional \ can be expanded as%
\begin{equation*}
\mathbf{f}=\frac{1}{\mathbf{v}^{2}}\left( \mathbf{f\cdot v}\right) \mathbf{v}%
+\mathbf{f}_{\bot },\quad \mathbf{f}_{\bot }=\frac{1}{\mathbf{v}^{2}}\left( 
\mathbf{v}^{2}\mathbf{f-v}\left( \mathbf{f\cdot v}\right) \right) ,\quad 
\mathbf{f}_{\bot }\cdot \mathbf{v=0.}
\end{equation*}%
We can turn the axis so that $\mathbf{v}$ \ is along $x_{3}$ and $\mathbf{f}%
_{\bot }$is \ along $x_{1}$, the sphere is preserved,\ and the norms of the
functionals $\mathbf{f}_{\bot }$ \ and $\mathbf{f-f}_{\bot }$ \ are
preserved. According to (\ref{xinej}) \ $\Xi \left( \mathbf{v},\mathbf{f}%
_{\bot }\right) =0$, therefore 
\begin{equation*}
\Xi \left( \mathbf{f},\mathbf{f}\right) =\Xi \left( \mathbf{f-f}_{\bot },%
\mathbf{f-f}_{\bot }\right) +\Xi \left( \mathbf{f}_{\bot },\mathbf{f}_{\bot
}\right) ,
\end{equation*}%
and the quadratic form \ takes the form 
\begin{equation*}
\Xi \left( \mathbf{f},\mathbf{f}\right) =\frac{1}{\mathbf{v}^{2}}\left( 
\mathbf{f\cdot v}\right) ^{2}\Xi \left( \frac{\mathbf{v}}{\left\vert \mathbf{%
v}\right\vert },\frac{\mathbf{v}}{\left\vert \mathbf{v}\right\vert }\right)
+\left\vert \mathbf{f}_{\bot }\right\vert ^{2}\Xi \left( \frac{\mathbf{f}%
_{\bot }}{\left\vert \mathbf{f}_{\bot }\right\vert },\frac{\mathbf{f}_{\bot }%
}{\left\vert \mathbf{f}_{\bot }\right\vert }\right) .
\end{equation*}%
Therefore 
\begin{equation*}
\Xi \left( \mathbf{f},\mathbf{f}\right) =\frac{1}{\mathbf{v}^{2}}\left( 
\mathbf{f\cdot v}\right) ^{2}\Xi _{\Vert }+\left\vert \mathbf{f}_{\bot
}\right\vert ^{2}\Xi _{\bot }.
\end{equation*}%
In particular, for a functional $\mathbf{f}=x_{i}$ 
\begin{equation}
\Xi \left( x_{i},x_{i}\right) =\frac{1}{\mathbf{v}^{2}}v_{i}^{2}\Xi _{\Vert
}+\left( 1-\frac{1}{\mathbf{v}^{2}}v_{i}^{2}\right) \Xi _{\bot }.
\label{ksiii}
\end{equation}%
For $i\neq j$ $\ \ \left\vert \left( x_{i}+x_{j}\right) \right\vert
^{2}=\left\vert \left( 1,1,0\right) \right\vert ^{2}=2$ and 
\begin{equation*}
\Xi \left( x_{i}+x_{j},x_{i}+x_{j}\right) =\frac{1}{\mathbf{v}^{2}}\left(
v_{i}+v_{j}\right) ^{2}\Xi _{\Vert }+\left( 2-\frac{1}{\mathbf{v}^{2}}\left(
v_{i}+v_{j}\right) ^{2}\right) \Xi _{\bot }.
\end{equation*}%
Therefore, for $i\neq j$ \ 
\begin{gather*}
\Xi \left( x_{i},x_{j}\right) =\frac{1}{4}\left( \Xi \left(
x_{i}+x_{j},x_{i}+x_{j}\right) -\Xi \left( x_{i}-x_{j},x_{i}-x_{j}\right)
\right) \\
=\frac{1}{4\mathbf{v}^{2}}\left[ \left( \left( v_{i}+v_{j}\right)
^{2}-\left( v_{i}-v_{j}\right) ^{2}\right) \Xi _{\Vert }+\left( \left(
v_{i}-v_{j}\right) ^{2}-\left( v_{i}+v_{j}\right) ^{2}\right) \Xi _{\bot }%
\right] ,
\end{gather*}%
hence%
\begin{equation*}
\Xi \left( x_{i},x_{j}\right) =\frac{v_{i}v_{j}}{\mathbf{v}^{2}}\left[ \Xi
_{\Vert }-\Xi _{\bot }\right] \text{ \ \ for \ }i\neq j.
\end{equation*}%
Since (\ref{ksiii}) can be written in the form 
\begin{equation*}
\Xi \left( x_{i},x_{i}\right) =\frac{v_{i}^{2}}{\mathbf{v}^{2}}\left[ \Xi
_{\Vert }\left( \left\vert \mathbf{v}\right\vert \right) -\Xi _{\bot }\left(
\left\vert \mathbf{v}\right\vert \right) \right] +\Xi _{\bot }\left(
\left\vert \mathbf{v}\right\vert \right) ,
\end{equation*}%
we obtain expression (\ref{ksid}) for the general case.


\begin{thebibliography}{99}
\bibitem{Appel Kiessling} Appel W. and Kiessling M., \textsl{Mass and Spin
Renormalization in Lorentz Electrodynamics}, Ann. Phys., \textbf{289},
24--83, (2001).

\bibitem{BF8} Babin A. and Figotin A., \textsl{Relativistic Dynamics of
Accelerating Particles Derived from Field Equations}, Found. Phys., \textbf{%
42}: 996--1014, (2012).

\bibitem{BF9} Babin A. and Figotin A., \textsl{Relativistic Point Dynamics
and Einstein Formula as a Property of Localized Solutions of a Nonlinear
Klein-Gordon Equation}, Comm. Math. Phys., 322 (2013), 453--499.

\bibitem{BF5} Babin A. and Figotin A., \textsl{Wave-corpuscle mechanics for
electric charges}, J. Stat. Phys., \textbf{138}: 912--954, (2010).

\bibitem{BF6} Babin A. and Figotin A., \textsl{Some mathematical problems in
a neoclassical theory of electric charges}, DCDS - Series A, \textbf{27}(4),
1283-1326 (2010).

\bibitem{BF7} Babin A. and Figotin A., Found. Phys., \textsl{Electrodynamics
of balanced charges}, \textbf{41}: 242--260, (2011).

\bibitem{BF10} Babin A. and Figotin A., \textsl{Newton's law for a
trajectory of concentration of solutions to nonlinear Schrodinger equation},
Commun. Pure Appl. Anal. 13 (2014), no. 5, 1685--1718. e-print available
online at arXiv:mathematics/1312.3688v1.

\bibitem{Barut} Barut A., \textsl{Electrodynamics of and Classical Theory of
Fields and Particles}, Dover, 1980.

\bibitem{CazenaveHaraux80} Cazenave T. and Haraux A., \textsl{\'{E}quations
d'\'{e}volution avec non lin\'{e}arit\'{e} logarithmique}, Ann. Fac. Sci.
Toulouse Math., \textbf{2} (1980), 21--51.

\bibitem{Dirac CE} Dirac P., \textsl{Classical Theory of Radiating Electrons}%
, Proc. Royal Soc., Series A, Vol. \textbf{167}, 929: 148-169, 1938.

\bibitem{Jackson} Jackson J., \textsl{Classical Electrodynamics}, 3rd
Edition, Wiley, 1999.

\bibitem{Kiessling 1} Kiessling M., \textsl{Electromagnetic Field Theory
without Divergence Problems 1. The Born Legacy}, J. Stat. Physics, \textbf{%
116}: 1057-1122, (2004).

\bibitem{Kiessling2} Kiessling M., \ \textsl{Quantum Abraham models with de
Broglie-Bohm laws of quantum motion}, in \textquotedblleft Quantum
Mechanics\textquotedblright\ (Festschrift for G.C. Ghirardi's 70th
birthday), A. Bassi, D. Duerr, T. Weber and N. Zanghi (Eds.), AIP Conference
Proceedings v.844, pp. 206--227 (2006), e-print available online at
arXiv:physics/0604069v2, (2007).

\bibitem{Kosyakov} Kosyakov B., \textsl{Introduction to the Classical Theory
of Particles and Fields}, Springer, 2007.

\bibitem{Lanczos VPM} Lanczos C., \textsl{The Variational Principles of
Mechanics}, 4th ed., Dover, 1986.

\bibitem{LandauLif F} Landau L. and Lifshitz E., \textsl{The classical
theory of fields, }Butterworth-Heinemann, Oxford, 1996.

\bibitem{LongS08} E. Long and D. Stuart, \textsl{Effective dynamics for
solitons in the nonlinear Klein-Gordon-Maxwell system and the Lorentz force
law}, Rev. Math. Phys., 21, 459--510 (2009).

\bibitem{Nodvik} Nodvik,\textsl{\ }J., \textsl{A covariant formulation of
classical electrodynamics for charges of finite extension, }Ann. Phys., 
\textbf{28}: 225, (1964).

\bibitem{Panofsky Phillips} Panofsky W. and Phillips M., \textsl{Classical
Electricity and Magnetism}, 2nd ed, Addison-Wesley, 1962.

\bibitem{Pearle1} Pearle P., in\textsl{\ Electromagnetism Paths to Research}%
, D. Teplitz, ed. (Plenum, New York, 1982), pp. 211--295.

\bibitem{Poincare} Poincar\'{e} H., Comptes Rend. 140, 1504 (1905);
Rendiconti del Circolo Matematico di Palermo, \textbf{21}, 129--176, (1906).

\bibitem{Rohrlich} Rohrlich F., \textsl{Classical Charged Particles},
Addison-Wesley, 3d ed., 2007.

\bibitem{Schwinger} Schwinger J.,\textsl{\ } \textsl{\ Electromagnetic Mass
Revisited}, Foundations of Physics, Vol. 13, No. 3, pp. 373-383, (1983).

\bibitem{Spohn} Spohn H., \textsl{Dynamics of Charged Particles and Their
Radiation Field}, Cambridge Univ. Press, 2004.

\bibitem{Thirring} Thirring W., \textsl{Classical Mathematical Physics,
Dynamical systems and field theory}, NY, Springer, 1997.

\bibitem{Wheeler Feynman 1} Wheeler J. and Feynman R.,\textsl{\ } \textsl{\
Interaction with the Absorber as the Mechanism of Radiation}, Rev. Mod.
Phys., Vol. 17, 2 and 3, 157-181, 1945.

\bibitem{Wheeler Feynman 2} Wheeler J. and Feynman R., \textsl{Classical
Electrodynamics in Terms of Direct Interparticle Action}, Rev. Mod. Phys.,
Vol. 21, 425-433, 1949.

\bibitem{Yaghjian} Yaghjian A., \textsl{Relativistic Dynamics of a Charged
Sphere: Updating the Lorentz-Abraham Model}, Springer, 2nd Ed., 2006.

\bibitem{Yaghjian1} Yaghjian A., \textsl{Absence of a consistent classical
equation of motion for a mass-renormalized point charge}, Phys. Rev. E. 
\textbf{78}, 046606, (2008).
\end{thebibliography}
\end{document}